\pdfoutput=1
\documentclass[fleqn,usenatbib]{mnras}
\usepackage{newtxtext,newtxmath}

\usepackage[T1]{fontenc}
\usepackage{ae,aecompl}

\usepackage{geometry}                		
\usepackage{graphicx}				
\usepackage{amssymb}

\usepackage{natbib}
\usepackage[usenames, dvipsnames]{color} 
\usepackage{cancel}
\usepackage{multicolumn}

\makeatletter
\def\@to{to}
\makeatother

\newcommand{\msun}{M$\rm{_{\odot}}$ }
\newcommand{\msunperiod}{M$\rm{_{\odot}}$}

\newcommand{\rsun}{R$\rm{_{\odot}}$}

\title{Signatures of Circumstellar Interaction in the Type IIL Supernova ASASSN-15oz}
\author[K. A. Bostroem et al.]{K. Azalee Bostroem, $\rm{^{1}}$\thanks{E-mail: kabostroem@ucdavis.edu}
Stefano Valenti,$\rm{^{1}}$
Assaf Horesh,$\rm{^{2}}$
\newauthor 
Viktoriya Morozova,$\rm{^{3}}$
N. Paul M. Kuin,$\rm{^{4}}$
Samuel Wyatt,$\rm{^{5}}$
\newauthor
Anders Jerkstrand,$\rm{^{6}}$
David J. Sand,$\rm{^{5}}$
Michael Lundquist,$\rm{^{5}}$
\newauthor
Mathew Smith,$\rm{^{7}}$
Mark Sullivan,$\rm{^{7}}$
Griffin Hosseinzadeh,$\rm{^{8}}$
Iair Arcavi,$\rm{^{9}}$
\newauthor
Emma Callis,$\rm{^{10}}$
R\'egis Cartier,$\rm{^{11}}$
Avishay Gal-Yam,$\rm{^{12}}$
Llu\'is Galbany,$\rm{^{13}}$
\newauthor
Claudia Guti\'errez,$\rm{^{7}}$
D. Andrew Howell,$\rm{^{14,\,15}}$
Cosimo Inserra,$\rm{^{17}}$
\newauthor
Erkki Kankare,$\rm{^{16}}$
Kristhell Marisol L\'opez,$\rm{^{18,\,19}}$
Curtis McCully,$\rm{^{14,\,15}}$
\newauthor
Giuliano Pignata,$\rm{^{20,\,21}}$
Anthony L. Piro,$\rm{^{22}}$
\'Osmar Rodr\'iguez,$\rm{^{20,\,21}}$
\newauthor
Stephen J. Smartt,$\rm{^{16}}$
Kenneth W. Smith,$\rm{^{16}}$
Ofer Yaron,$\rm{^{12}}$
\newauthor
David R. Young$\rm{^{16}}$
\\
$\rm{^{1}}$Department of Physics, University of California, Davis, CA 95616, USA \\
$\rm{^{2}}$Racah Institute of Physics, The Hebrew University of Jerusalem, Jerusalem, 91904, Israel \\ 
$\rm{^{3}}$Department of Astrophysical Sciences, Princeton University, Princeton, NJ 08544, USA \\ 
$\rm{^{4}}$Mullard Space Science Laboratory, University College London, Holmbury St.~Mary, Dorking, Surrey RH5 6NT, UK \\ 
$\rm{^{5}}$Department of Astronomy/Steward Observatory, 933 North Cherry Avenue, Rm. N204, Tucson, AZ 85721-0065, USA \\ 
$\rm{^{6}}$Max-Planck Institut f{\"u}r Astrophysik, Karl-Schwarzschild Str. 1, D-85748 Garching, Germany \\ 
$\rm{^{7}}$School of Physics and Astronomy, University of Southampton, Southampton, SO17 1BJ, UK\\ 
$\rm{^{8}}$Harvard-Smithsonian Center for Astrohysics,  60 Garden Street,  Cambridge, MA 02138 \\ 
$\rm{^{9}}$The Raymond and Beverly Sackler School of Physics and Astronomy, Tel Aviv University, Tel Aviv 69978, Israel \\ 
$\rm{^{10}}$School of Physics, O'Brien Centre for Science North, University College Dublin, Belfield, Dublin 4, Ireland \\ 
$\rm{^{11}}$Cerro Tololo Inter-American Observatory, National Optical Astronomy Observatory, Casilla 603, La Serena, Chile \\ 
$\rm{^{12}}$Department of Particle Physics and Astrophysics, Weizmann Institute of Science, Rehovot, 7610001, Israel \\ 
$\rm{^{13}}$PITT PACC, Department of Physics and Astronomy, University of Pittsburgh, Pittsburgh, PA 15260, USA\\
$\rm{^{14}}$Department of Physics, University of California, Santa Barbara, CA 93106-9530, USA\\ 
$\rm{^{15}}$Las Cumbres Observatory, 6740 Cortona Dr Ste 102, Goleta, CA 93117-5575, USA\\ 
$\rm{^{16}}$ Astrophysics Research Centre, School of Mathematics and Physics, Queens University Belfast, Belfast BT7 1NN, UK \\ 
$\rm{^{17}}$School of Physics \& Astronomy, Cardiff University, Queens Buildings, The Parade, Cardiff, CF24 3AA, UK\\ 
$\rm{^{18}}$ SRON Netherlands Institute for Space Research, Sorbonnelaan 2 3584 CA Utrecht, The Netherlands\\ 
$\rm{^{19}}$ Department of Astrophysics/IMAPP, Radboud University, P.O. Box 9010, 6500 GL Nijmegen, The Netherlands\\ 
$\rm{^{20}}$Departamento de Ciencias Fisicas, Universidad Andres Bello, Avda. Republica 252, Santiago, Chile \\ 
$\rm{^{21}}$Millennium Institute of Astrophysics (MAS), Nuncio Monse{\~n}or S{\'o}tero Sanz 100, Providencia, Santiago, Chile \\ 
$\rm{^{22}}$ The Observatories of the Carnegie Institution for Science, 813 Santa Barbara Street, Pasadena, CA 91101, USA\\ 
}
\date{Accepted XXX. Received YYY; in original form ZZZ}
\pubyear{2018}
\begin{document}
\label{firstpage}
\pagerange{\pageref{firstpage}--\pageref{lastpage}}
\maketitle
\begin{abstract}
Hydrogen-rich, core-collapse supernovae are typically divided into four classes: IIP, IIL, IIn, and IIb. In general, interaction with circumstellar material is only considered for Type IIn supernovae. However, recent hydrodynamic modeling of IIP and IIL supernovae requires circumstellar material to reproduce their early light curves. In this scenario, IIL supernovae experience large amounts of mass loss before exploding.
We test this hypothesis on ASASSN-15oz, a Type IIL supernova. With extensive follow-up in the X-ray, UV, optical, IR, and radio we present our search for signs of interaction, and the mass-loss history indicated by their detection. We find evidence of short-lived intense mass-loss just prior to explosion from light curve modeling, amounting in 1.5 M$\rm{_{\odot}}$ of material within 1800 R$\rm{_{\odot}}$ of the progenitor. We also detect the supernova in the radio, indicating mass-loss rates of $\rm{10^{-6}}$ -10$\rm{^{-7}}$ M$\rm{_{\odot}}$ yr$\rm{^{-1}}$ prior to the extreme mass-loss period. Our failure to detect the supernova in the X-ray and the lack of narrow emission lines in the UV, optical, and NIR do not contradict this picture and place an upper limit on the mass-loss rate outside the extreme period of $\rm{<10^{-4}}$ M$\rm{_{\odot}}$yr$\rm{^{-1}}$. 
This paper highlights the importance gathering comprehensive data on more Type II supernovae to enable detailed modeling of the progenitor and supernova which can elucidate their mass-loss histories and envelope structures and thus inform stellar evolution models.
\end{abstract}
\section{Introduction}
Stars greater than 8 \msun end their lives spectacularly as core-collapse supernovae (CCSN). 
CCSNe are divided into two types: Type I, those without hydrogen in their spectra and Type II, those with hydrogen in their early spectra \citep{1941minkowski}.
Although the diversity of Type II light curves was noted early \citep{1964minkowski, 1967pskovskii}, it was \citet{1979barbon} who first proposed a division of the Type II class into Type IIP SNe (IIP SNe) and Type IIL SNe (IIL SNe) based on the shape of their light curves.
With small samples, some authors have found a distinction between IIP SNe and IIL SNe light curves  (e.g. \citealt{1993patat}, and \citealt{1994patat}, \citealt{2012arcavi}).
However, the recent analysis of large samples of SNe by \citet{2014anderson, 2014faran,2015sanders,2016valenti,2016galbany}, and \citet{2016rubin} show a continuum between the archetypical IIP and IIL light curves. 
Given the continuum of light curves observed, we will use IIP-like to denote SNe at the IIP end of the spectrum and IIL-like to denote SNe at the IIL end of the spectrum.

From high resolution imaging prior to the SNe explosion, the progenitors of IIP/IIL SNe are found to be red supergiants (RSG) with masses $\rm{<}$ 17  \msun (\citealt{2015smartt} and references there in).
Successful modeling of IIP-like SNe with RSG progenitors has shown that the initial period of decline is due to adiabatic cooling during the expansion of the ejecta (e.g. \citealt{1971grassberg,1977falk,1993blinnikov,2009kasen}).
When the photosphere reaches $\rm{\sim}$ 6000K, hydrogen recombination commences, moving the photosphere inward, against the outwardly expanding ejecta. 
The rate of the expansion of the ejecta matches the rate of the recession of the recombination front, which itself is at the constant temperature of recombination, leading to the almost constant brightness that define the IIP-like class.
When the photosphere passes completely through the hydrogen dominated layer of the ejecta, the ejecta continues to expand and cool, as layers rapidly become optically thin until the light curve is powered by the radioactive decay of $\rm{{}^{56}Ni \rightarrow {}^{56}Co \rightarrow {}^{56}Fe}$.

It is theorized that IIL-like SN light curves arise from progenitors with  smaller hydrogen envelopes \citep{1971grassberg,1989branch,1993blinnikov}.
In this scenario, the recombination front would move faster than the expanding ejecta, causing the brightness to decrease. 
The shallow H $\rm{\alpha}$ absorption troughs in IIL-like SNe, possibly due to less absorbing material, support this picture \citep{1996schlegel,2014gutierrez}.
To explode with smaller hydrogen envelopes, IIL-like SN progenitors must lose mass via stellar winds or mass transfer to a binary companion.
In this paper we will consider the case of mass loss via stellar winds.
As mass-loss increases with zero age main sequence mass (M$\rm{_{ZAMS}}$)  \citep{2003heger,2009kasen}, this leads to the hypothesis that the progenitors of IIL-like SNe are more massive than IIP-like progenitors. 
Observations that IIL-like SNe have shorter periods before falling to the radioactive decay phase, are brighter, and have higher velocity ejecta support this hypothesis \citep{1994patat,2014gutierrez,2014anderson,2014faran,2015sanders,2015valenti,2016valenti}.
This idea is intriguing as it solves another outstanding problem, the lack of progenitors between 17-25 \msun \citep{2015smartt} (for other possible solutions see: \citealt{2018davies, 2012walmswell}).
However, \citet{2016valenti} found that there is not a clear difference in the progenitor population of IIP-like and IIL-like SNe and more importantly, that the progenitors of IIL-like SNe are not above 17 \msunperiod.
Nevertheless, mass-loss, whether governed by progenitor mass or some other factor (e.g. metallicity, rotation, binarity) remains the most promising explanation for the diversity of light curve shapes.

Mass-loss rates for RSGs are not well understood theoretically and are extremely difficult to characterize with observations due to both the short period of this phase as well as the rarity of high mass stars.
Thus the mass loss rates for RSG are uncertain, with typical mass-loss rate estimates between $\rm{10^{-6}-10^{-4}}$ \msunperiod $\rm{yr^{-1}}$ and wind velocities between $\rm{10-100}$ km $\rm{s^{-1}}$ \citep{2011mauron}.
Mass lost from a SN progenitor over the course of its lifetime will be distributed around the progenitor with a density that is related to the time the mass-loss occurred, the mass-loss rate, and the wind velocity.
Although interaction has been observed in SNe at the edge of the distribution in terms of decline rate (steep) and absolute magnitude (bright; e.g. SN 1979C: \citealt{1993blinnikov}; SN 1980K: \citealt{1992chugai}; SN 1998S: \citealt{2001chugai}; see Section \ref{SecComp}), it is often assumed to be negligible in IIP/IIL SNe.
It is possible that the signatures of interaction with circumstellar material (CSM; with stellar wind origins) are visible in the light curves and spectra of most IIP/IIL SNe, especially outside of the optical where there are fewer observations.
The detection and evolution of this interaction can directly probe the mass-loss history of massive stars and provide valuable information on the nature of IIP/IIL SN progenitors.

In fact, evidence of interaction, often characterized by a single type of observation (e.g. X-ray detection), indicates that mass-loss should be considered in the interpretation of IIP/IIL SNe.
Recent light curve modeling has shown that dense CSM is required to match the early UV and optical light curves of IIP/IIL SNe with hydrodynamic models of SNe from RSG progenitors \citep{2015gezari,2018morozova,2018paxton,2018foerster}.  
Flash ionization features, the narrow emission lines in Type IIP/IIL SNe that disappear within the first hours to days post explosion are also explained by interaction with CSM. 
These features can be produced either by episodic high mass-loss rates in single stars (\citealt{2014gal-yam}, e.g. \citealt{ 2017yaron}) or by lower mass-loss rates in a binary system \citep{2018kochanek}.
Additional evidence of the importance of mass-loss comes from rare X-ray and radio observations of IIP/IIL SNe, some of which detect signs of interaction (e.g. \citealt{2014dwarkadas} and references there in, \citealt{2018morozova2}).

Although interaction is seen in some historical IIL-like SNe (e.g. 1979C,1980K, 1998S), as more IIL-like SNe are discovered, it has become clear that these interacting events are at the edge of the IIL-like distribution with bright absolute magnitudes and steep decline rates. 
Furthermore, a large number of IIL-like SNe, more central to the distribution, lack strong signs of interaction.
Here, we search for interaction in one of these central IIL-like SN, discovered by the All-Sky Automated Survey for SuperNovae (ASAS-SN), ASASSN-15oz.
With our extensive set of data, ranging from X-ray to radio wavelengths, we examine different indicators of interaction with CSM to form a coherent history of mass-loss.

The outline of the paper is as follows.
In Section \ref{sec:Obs} we describe the data collected at all wavelengths.
In Section \ref{15ozIntro} we present the parameters of ASASSN-15oz.
The photometric and spectroscopic evolution are described in Sections \ref{sec:LCEvolve} and \ref{sec:SpecEvolve}, respectively.
Section \ref{sec:Interaction} describes our search for interaction, including hydrodynamic light curve modeling and radio analysis.
Finally, we present a comparison to other objects and a unified interpretation of the mass-loss history indicated by the multi-wavelength observations  in Section \ref{SecComp} and summarize our conclusions in Section \ref{sec:conclude}.
\section{Observations and Data Reduction}  \label{sec:Obs}
ASASSN-15oz was the closest Type II SN of 2015 and was observed extensively at all wavelengths. 
The majority of our observations are provided by the Las Cumbres Observatory (LCO; \citealt{2013brown}) Supernova Key Project (2014-2017) and the Public ESO Spectroscopic Survey for Transient Objects (PESSTO). 
We supplemented these with observations in the X-ray, ultraviolet (UV), optical, near infra-red (NIR), and radio. 
The photometric and spectroscopic data reduction is described in this section. 
A table of spectroscopic observations is presented in Table \ref{tab:SpecObs}.
A complete list of photometric observations is given in the Supplemental Material. 
Table \ref{tab:LcObs} gives an example of the photometric observation table format.  
All photometric data can be obtained from the SNDAVIS database\footnote{http://dark.physics.ucdavis.edu/sndavis/transient} and all spectroscopic data can be found on WISeREP\footnote{https://wiserep.weizmann.ac.il}.
\subsection{Optical \label{subsec:optical}}
\begin{figure}
\begin{center}
\includegraphics[width=\columnwidth]{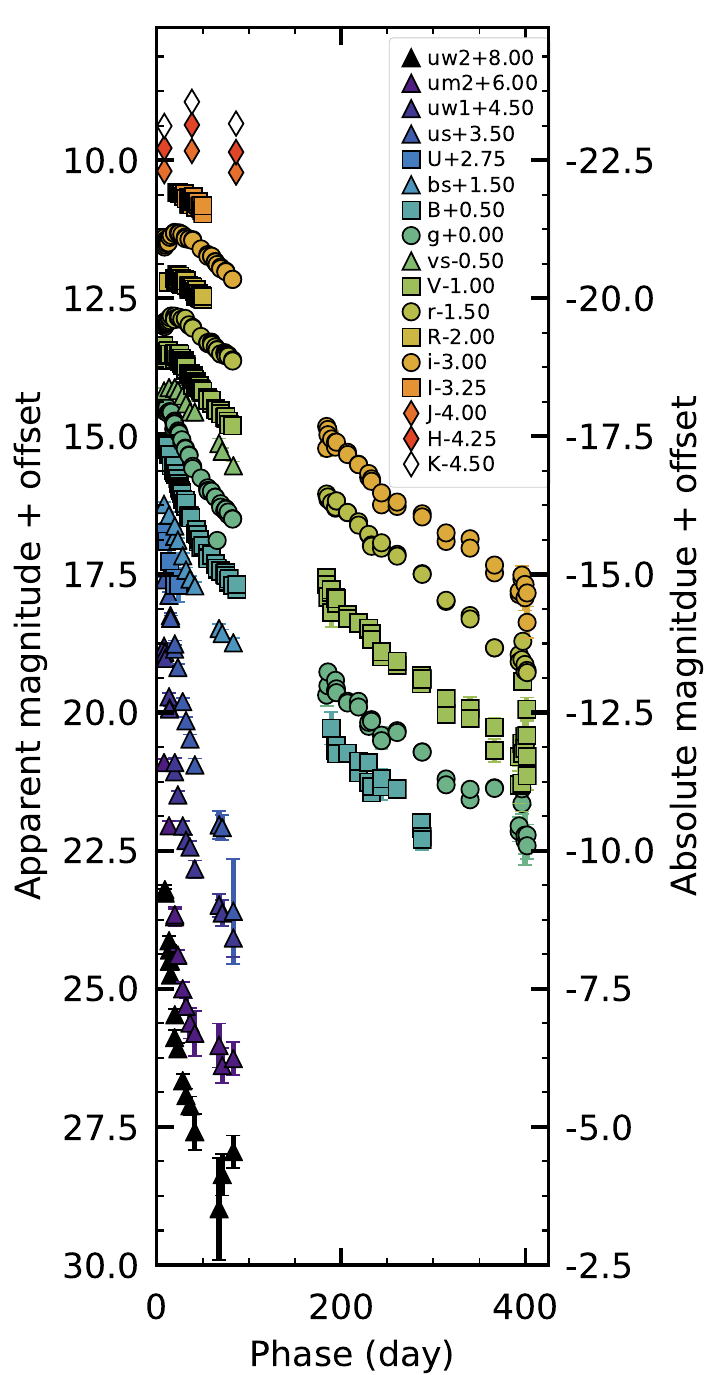}
\caption{The complete UV, optical, and NIR light curve of ASASSN-15oz expressed in terms of days since explosion.
Apparent magnitude (+offset) is plotted on the left and absolute magnitude (+offset) is plotted on the right. 
Each color corresponds to a single filter, labeled in the legend. 
The optical {\it Swift} filters are denoted with an "s" in the name (e.g. us) and plotted with the triangle symbol.
NIR observations are shown with the diamond symbol.
To differentiate the {\it U, B, V, R,} and {\it I} filters and the {\it g, r,} and {\it i} filters, we use a square symbol for the former and a circle for the later.
Each filter has been shifted by a global offset (denoted in the legend) for viewing purposes.}
\label{fig:LC}
\end{center}
\end{figure}

The optical light curve was closely monitored by the LCO from discovery through day 400 in the filters {\it B, V, g, r,} and $i$. 
Unfortunately, the SN passed behind the sun around day 80, leaving a gap in the observations from day 87 to day 179, most notably missing the fall from plateau.
Figure \ref{fig:LC} shows the complete multi-band light curve.
LCO observations were reduced with the LCO imaging pipeline, {\tt lcogtsnpipe} \citep{2016valenti}. 
This pipeline employs PSF photometry, removing background contamination by fitting a low order polynomial to the host galaxy.
Instrumental magnitudes were converted to apparent magnitudes using stars in the the APASS\footnote{https://www.aavso.org/apass}, Sloan, and Landolt catalogs.
Apparent magnitudes for the {\it U, B, V, R,} and {\it I} filters are given in the Vega magnitude system and apparent magnitudes for  the {\it g, r,} and {\it i} filters are given in the AB magnitude system.

Extensive optical spectroscopy was obtained during the photospheric phase with the FLOYDS spectrograph \citep{2013brown} on the 2m LCO telescope at Siding Springs, Australia and the EFOSC2 spectrograph \citep{1984buzzoni} on the 3.6 m NTT telescope at La Silla, Chile during the photospheric and nebular phase. 
One-dimensional spectra were extracted and calibrated using the FLOYDS pipeline \citep{2014valenti} and the EFOSC tasks in the PESSTO pipeline \citep{2015smartt}. 
Both of these pipelines combine IRAF tasks to bias subtract and flat field the data and locate, extract, wavelength calibrate, and flux calibrate the one-dimensional spectrum.

Data were also taken using the X-Shooter echelle spectrograph \citep{2011vernet} on the 8.2m Very Large Telescope (VLT) at the European Southern Observatory (ESO).  
X-Shooter has three arms (UVB, VIS, NIR) which combined provide continuous wavelength coverage over $\rm{3000-24,800}$ \AA. 
 For our data, we used slit widths of 1.0, 0.9, and 0.9 arcsec in the UVB, VIS, and NIR arms, corresponding to resolutions of R $\rm{\sim}$5400, 8900, and 5600, respectively. 
The UVB and VIS data were reduced via a modified version of the {\tt EsoReflex} pipeline \citep{2013freudling}, with improvements to the sky-subtraction and rebinning procedure on the highly dispersed echelle spectrum.

\begin{figure*}
\begin{center}
\includegraphics[width=\textwidth]{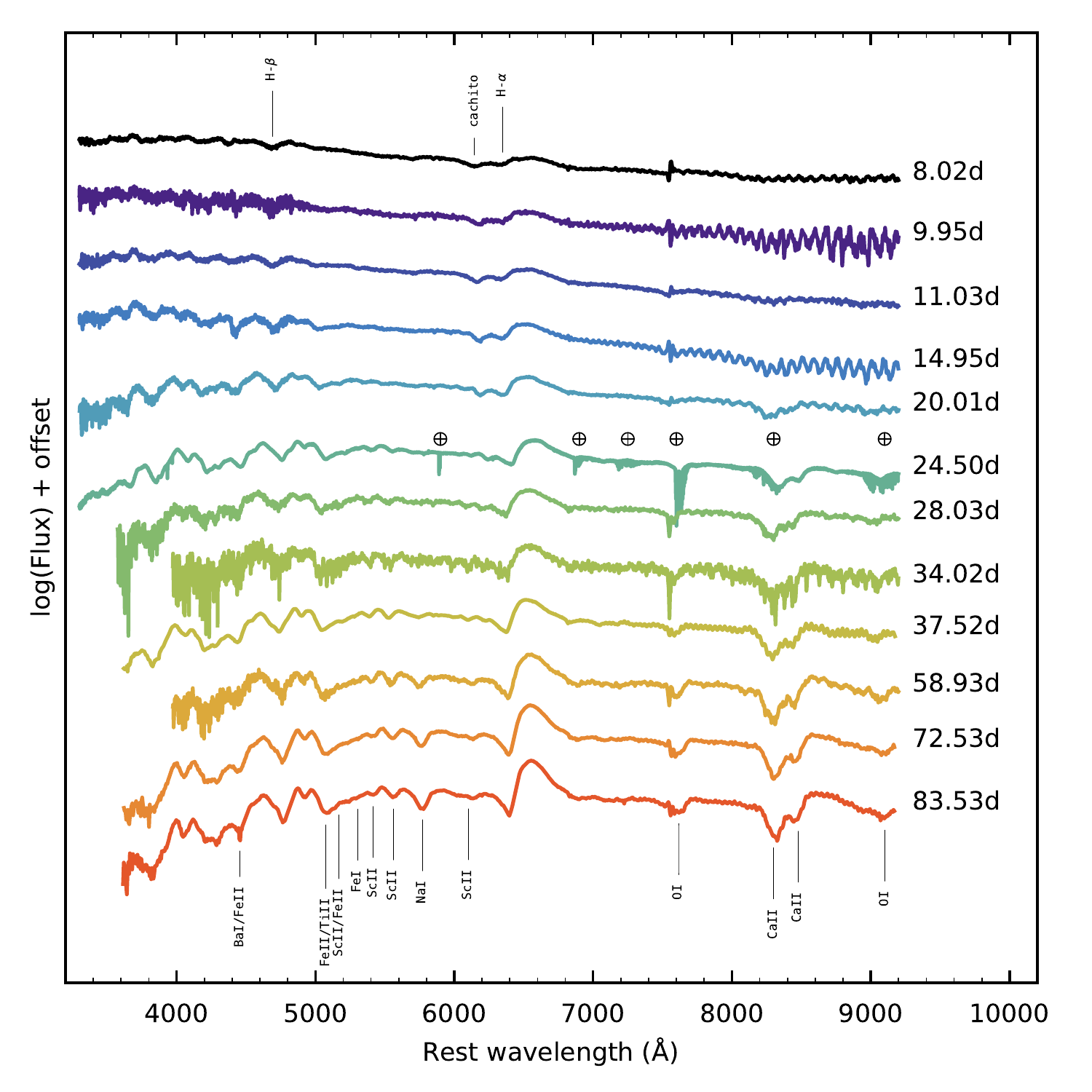}
\caption{A time series of the photospheric spectra of ASASSN-15oz at rest wavelengths.
The phase of each spectrum is marked on the right.
Potential telluric contamination is marked with crossed circles above the day 24.5 spectrum. 
The first spectrum taken near maximum light shows hydrogen features (identified at the top of the figure).
At this early phase, the cachito feature (see Section \ref{sec:cachito}) is already visible. 
Over time the cachito feature fades while the hydrogen emission grows and metal lines become visible and grow in strength. 
These lines are identified in the bottom spectrum.}
\label{fig:SpecAll}
\end{center}
\end{figure*}

Figure \ref{fig:SpecAll} shows the evolution of the optical spectra over time.
The progression starts with a blue spectrum with broad hydrogen lines at the top of the figure.
Over time (moving down the figure), the spectra develop metal features such as iron and scandium as the velocities decrease and the line widths narrow.

We obtained late time spectra from the Gemini South Observatory through program GS-2016A-Q-75-25 (PI: Valenti).  
The Gemini spectra of ASASSN-15oz were taken on day 287 (2016 Jun 09), day 288 (2016 Jun 10), and day 290 (2016 Jun12) using the Gemini Multi-Object Spectrograph (GMOS) \citep{2004hook}.  
Data were taken with a blue setting with the B600 grating, no filter, a 1.0\arcsec slit, and 2x2 binning, and a red setting with the R400 grating, OG515 order blocking filter, a 1.0\arcsec slit, and 2x2 binning.  
These settings provide a wavelength range 3630 - 6830 \AA \ with a dispersion of 1.0 \AA \ $\rm{pixel^{-1}}$ in the blue and a wavelength range 5030 - 9800 \AA \ with a dispersion of 1.3\AA \ $\rm{pixel^{-1}}$ in the red.
We used the {\tt Gemini iraf} package for GMOS to process our data.  
The raw two-dimensional images were bias subtracted, overscan corrected, flat-fielded, wavelength calibrated using CuAr emission lamps, cleaned of cosmic rays, and background subtracted prior to extraction of the 1D target spectra.  
The spectra were then calibrated using a sensitivity function derived from a standard star observed for the program.
\subsection{UV and X-Ray Observations}\label{SecSwift}
\begin{figure}
\begin{center}
\includegraphics[width=\columnwidth]{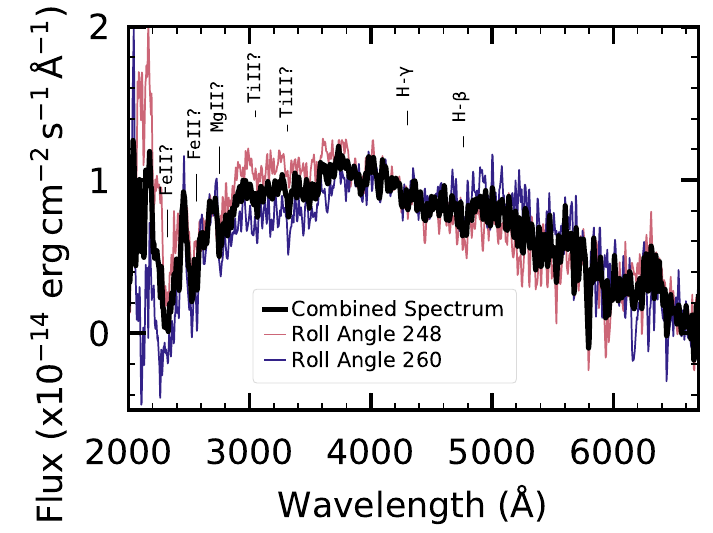} 
\caption{The {\it Swift }UV spectra from day nine (2015 Sep 05), boxcar smoothed with a kernel of three pixels. 
The spectrum taken with a roll angle of PA = 248$\rm{^{\circ}}$ is in pink, the  spectrum taken with a roll angle of PA = 260$\rm{^{\circ}}$ in indigo, and average flux is in black.
We tentatively identify UV FeII, TiII, and MgII features as well as mark the Balmer lines.
These spectra are not template subtracted, therefore they show residuals from zeroth order light. These features can be identified as being in the spectrum of one roll angle but not the other.
The lack of narrow line emission indicates a lack of strong interaction.}
\label{fig:SwiftSpectrum}
\end{center}
\end{figure}

UV and X-Ray imaging observations were obtained concurrently with \textit{Swift} from day nine (2015 Sep 05) through day 83 (2015 Nov 18).
In addition to the imaging observations, two epochs were observed with the \textit{Swift} grism under program 1114241 (PI: Valenti). 
Imaging observations were processed with aperture photometry following \citet{2009brown} and employing the updated zeropoints of \citet{2010breeveld}.
UV grism observations were taken at two roll angles and extracted using the UVOTPY pipeline\footnote{\citet{2014kuin}} \citep{2015kuin}. 
There is a nearby star coincident with the SN at one roll angle and just above it at the other roll angle.
Although we tried custom extractions using different extraction box height and template subtraction using TRUVOT \citep{2015smitka} we were unable to produce a better calibration than the default pipeline. 
Due to the distance of the ASASSN-15oz, only the first of the two spectroscopic epochs has sufficient S/N to be extracted. 
These spectra are shown in Figure \ref{fig:SwiftSpectrum}.

Aperture photometry was performed on the X-ray observations using the  {\tt HEAsoft} packages {\tt xselect} \citep{blackburn_1995} and {\tt xspec}; \citep{arnaud_1996}.
The background was selected as a region with no x-ray sources in the image.
The source was extracted using an 18\arcsec aperture.
Count rates were converted to a luminosity using PIMMS \citep{mukai_1993} assuming a power-law model with a photon index of two.
Integrating over all available SWIFT/XRT epochs, no significant detection was made after background subtraction.
\subsection{NIR Observations}
ASASSN-15oz was observed in the NIR with the SOFI instrument on the NTT telescope through the PESSTO collaboration, the SpeX instrument \citep{2003rayner} on the 3.0-m NASA Infrared Telescope Facility (IRTF), and the NIR arm of the X-Shooter spectrograph. 
The data reduction of the NIR arm of the X-Shooter spectrum is described with the UV and VIS arms in Section \ref{subsec:optical}.

The SOFI spectroscopic data were reduced with the PESSTO package \citep{2015smartt}.  
Photometric measurements were performed with the QUBA pipeline \citep{2011valenti}, which performs DAOPHOT-based \citep{1987stetson} point-spread-function fitting photometry on the SN and on the selected reference stars. 
{\it J, H, \rm{and} K} photometry was then calibrated to the 2MASS magnitude system.

The SpeX data were taken in so-called SXD mode, where the spectrum is cross-dispersed to obtain wavelength coverage from $\rm{\sim}$0.8--2.4 $\rm{\mu}$m in a single exposure, spread over six orders.  
All observations were taken with the slit aligned along the parallactic angle, and we employed a classic ABBA technique for improved sky subtraction.  
HD 177074, an A0V star, was observed adjacent in time to the science observations for flux and telluric calibration.  
The spectrum was reduced in a standard way using the publicly available {\sc Spextool} software package \citep{2004cushing} and corrections for telluric absorption utilized {\sc XTELLCOR} \citep{2003vacca} and the A0V star observations.

\begin{figure*}
\begin{center}
\includegraphics[width=\textwidth]{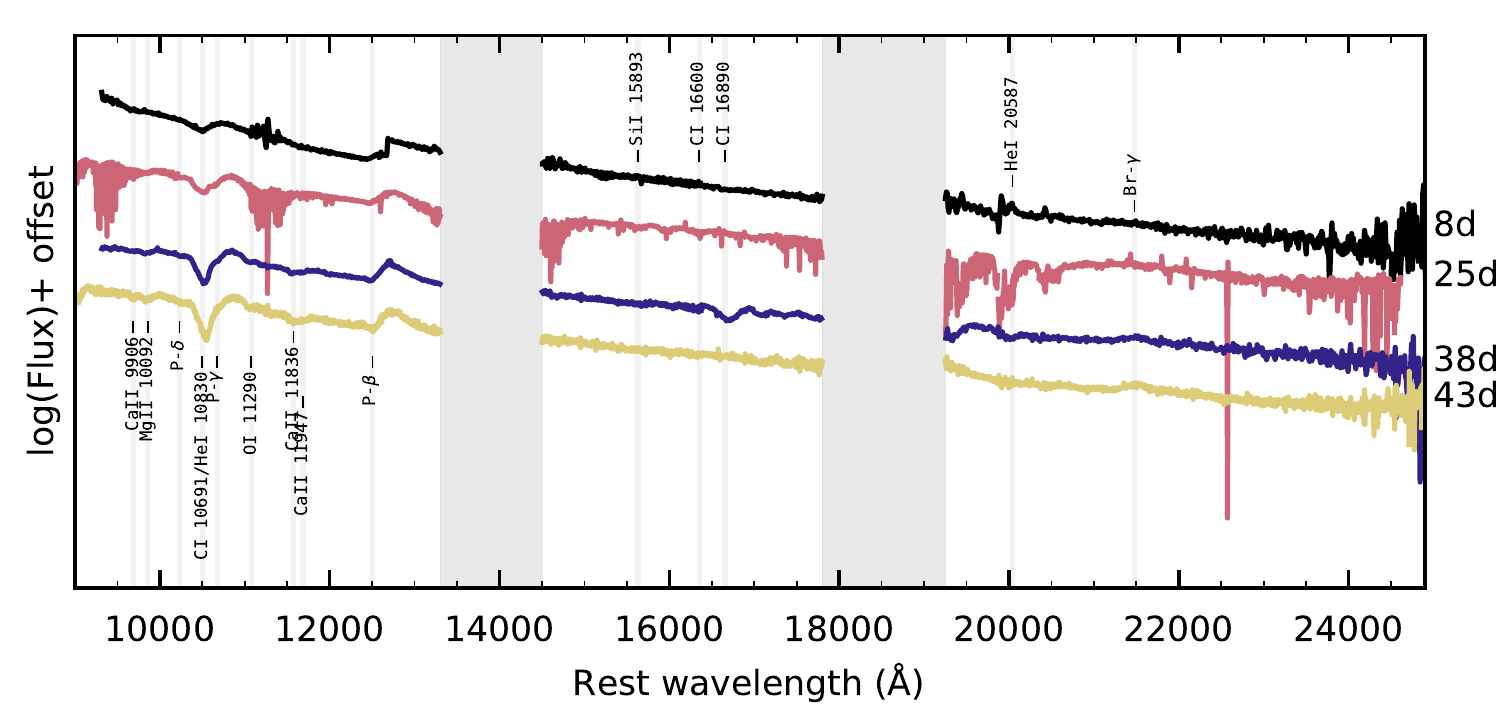}
\caption{The evolution of the infrared spectra at rest wavelengths during the photospheric phase. 
The spectral evolution in the IR mirrors the optical evolution, with broad hydrogen features narrowing and metal lines developing over time.
The first spectrum (black) taken eight days post-explosion shows broad hydrogen features.
The later spectra at day 38 (indigo) and 43 (yellow) reveal the development metal lines.
Although contaminated with narrow telluric features, the high resolution X-Shooter spectrum (pink) shows an intermediate phase.
Regions of severe sky contamination are marked with grey boxes and are masked out of the specta.}
\label{fig:IRmontage}
\end{center}
\end{figure*}

The three photometric observations from day eight (2015 Sep 04), day 38 (2015 Oct 04), and day 86 (2015 Nov 21) are shown in Figure \ref{fig:LC}.
Spectroscopic observations from SOFI (day nine; 2015 Sep 05 and day 39; 2015 Oct 05), X-Shooter (day 26; 2015 Sep 21), and SpeX (day 44; 2015 Oct 10) are shown in Figure \ref{fig:IRmontage}. 
\subsection{Radio Observations}
We observed ASASSN-15oz using the NSF's Karl G. Jansky Very Large Array (VLA) (VLA/15B-362, PI: Valenti) on day 28 (2015 Sep 24) and day 42 (2015 Oct 08). 
In both of the observations we used J1924-2914 and 3C48 as a phase calibrator and a flux calibrator, respectively. 
We used the Common Astronomy Software Applications (CASA) standard packages and pipelines to perform the data calibration and imaging. 
The SN was detected in both observation with the following flux measurements: September 24: $\rm{120\pm 23\,\mu}$Jy at $\rm{4.8}$\,GHz (C-band) and a $\rm{3\,\sigma}$ non-detection limit at $\rm{22}$\,GHz (K-band); October 08:  $\rm{210\pm 21\,\mu}$Jy at $\rm{4.8}$\,GHz (C-band) and a $\rm{80\pm 17\,\mu}$Jy at $\rm{15}$\,GHz (KU-band). 
\section{Supernova Parameters} \label{15ozIntro}
\begin{figure}
\begin{center}
\includegraphics[width=\columnwidth]{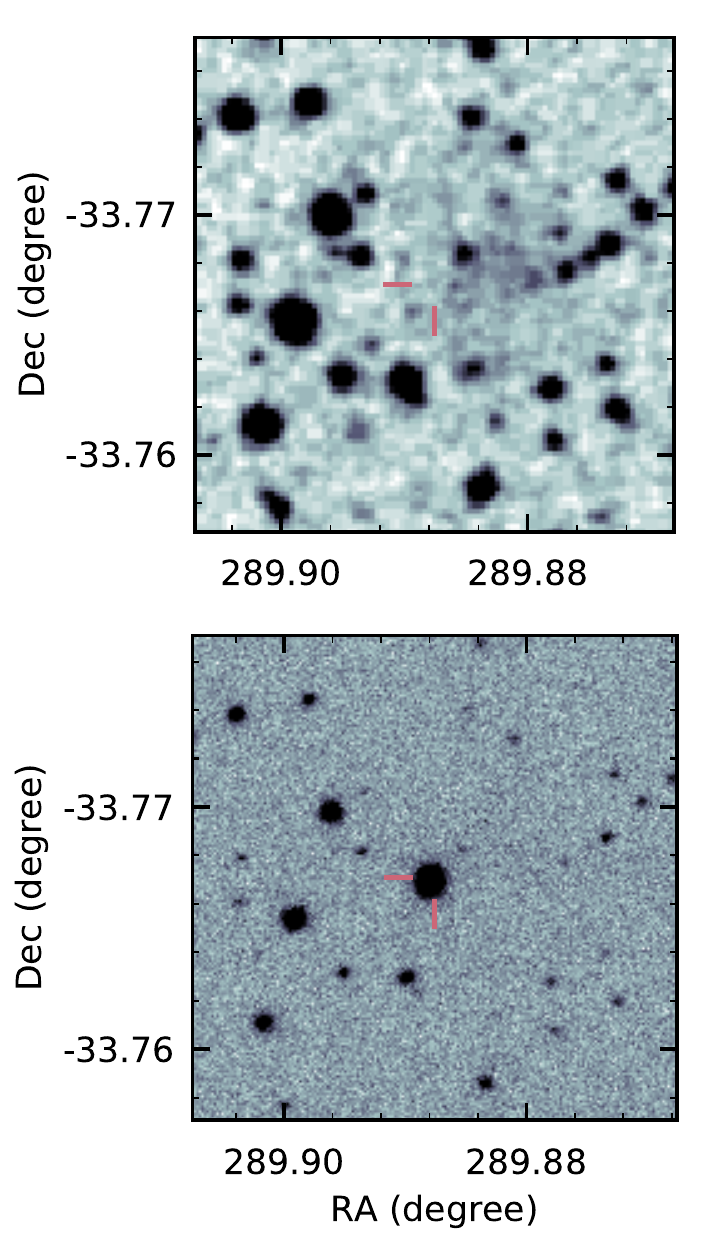} 
\caption{Top: The host galaxy of ASASSN-15oz, HIPASS J1919-33, taken prior to the SN explosion by DSS. 
Bottom: {\it V}-band image of ASASSN-15oz taken by LCO eight days post explosion (2015 Sep 04).
The location of the ASASSN-15oz is marked by the pink ticks in both images.}
\label{fig:finder}
\end{center}
\end{figure}

ASASSN-15oz was discovered by the ASAS-SN team on 2015 Aug 31.09 (JD 2457265.59; V = 14.6 mag; RA = 19:19:33.55, Dec=-33:46:01.2) and announced on 2015 Sep 03 (JD 2457268; \citealt{2015brown}). 
The next day, 2015 Sep 04 (JD 2457270.0), LCO classified it as a Type II SN one week after explosion \citep{2015hosseinzadeh}. 
Figure \ref{fig:finder} shows the Digital Sky Survey (DSS) image of the low luminosity host galaxy, HIPASS J1919-33, prior to explosion (top panel) and our first {\it V}-band image of ASASSN-15oz taken on day eight (2015 Sep 04).
Imaging of the host galaxy, HIPASS J1919-33, from 2015 Aug 23 (JD 2457257.56) does not show any evidence of ASASSN-15oz (V $\rm{>}$ 17.8; \citealt{2015brown}). 
We take this date to be a lower limit on the explosion epoch and define the mid-point between the non-detection and the detection as explosion date: $\rm{t_{expl} =}$ 2015 Aug 27 $\rm{\pm 4}$ days (JD 2457261.5). 
This is in agreement with the value used in \citet{2018gutierrez}.\\
\indent When non-detection observations are not available, other methods can be used to determine the explosion epoch. 
\citet{2017gutierrez} found that the explosion epoch can be determined, with an average errors of 4.9 days, by fitting the blue part of the first spectrum with SNID\footnote{\citet{2011blondin}} 
\citep{2007blondin} and using the explosion epoch of the best fit template.
This method relies on the existence of a template SN that is similar in both phase and type to the object being fit.
ASASSN-15oz is a IIL-like SN. 
SNID has very few templates for IIL-like SNe, even after adding the new templates provided by \citet{2017gutierrez}. 
Therefore, in addition to the \citet{2017gutierrez} templates, we add templates for two well observed supernovae with well constrained explosion epochs: SN 2012A \citep{2013tomasella}, and SN 2013ej \citep{2016childress,2016dhungana,2014valenti,2015smartt}.
With these templates, we confirm our explosion epoch by fitting the first spectrum (2015 Sep 04) using SNID v. 5.0.
The five best fit spectra imply an explosion epoch of  2015 Aug 22$\rm{\pm}$4.9 (JD 2457256.5), consistent with the range we propose based on non-detections.
Interestingly, the hydrodynamic light curve modeling also places the best fit explosion epoch earlier than 2016 Aug 27 (see Section \ref{sec:LCEvolve}).
As the uncertainties of both these methods are greater than that of the non-detection, throughout this paper we adopt the value determined by the non-detection as the explosion epoch, $\rm{t_{expl} =}$ 2015 Aug 27 $\rm{\pm 4}$ days (JD 2457261.5), unless otherwise specified.

\begin{figure}
\begin{center}
\includegraphics[width=\columnwidth]{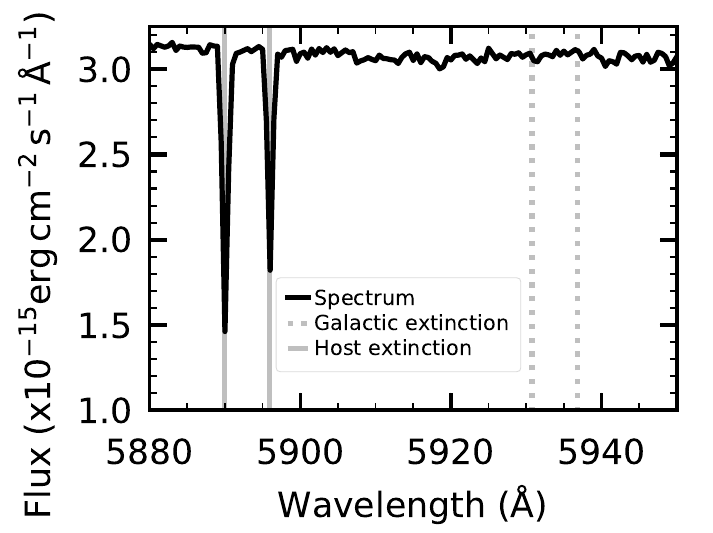} 
\caption{The galactic (solid gray lines) and host (dashed gray lines) Na I D1 and D2 lines in the high resolution X-shooter spectrum from day 26; (2015 Sep 21).
The galactic profiles are consistent with the galactic extinction of \citet{2011schlafly} while the absence of absorption at the host redshift indicates a lack of dust in the host galaxy. }
\label{fig:extinction}
\end{center}
\end{figure}

We use the Hubble flow distance modulus to the host galaxy, HIPASS J1919-33,  $\rm{\mu=32.3\pm 0.2}$ mag, with H $\rm{_{0}}$=73 km $\rm{s^{-1}}$ $\rm{Mpc^{-1} }$ corrected for Virgo infall
\footnote{The NASA/IPAC Extragalactic Database (NED) is operated by the Jet Propulsion Laboratory, California Institute of Technology, under contract with the National Aeronautics and Space Administration.}.
We use the VIS arm of the X-shooter spectrum (R $\rm{\sim}$8900) to constrain the host galaxy reddening and to confirm the galactic reddening using the equivalent widths of the Na I D1 and D2 lines ($\rm{\lambda}$5890, $\rm{\lambda}$5896) \citep{2012poznanski}. 
We find no evidence of Na I absorption at the host redshift and therefore assume no host reddening (see Figure \ref{fig:extinction}). 
By fitting gaussian profiles to the galactic Na I absorption we find $\rm{E(B-V) = 0.12\pm0.07}$ mag, consistent with the value given by \citet{2011schlafly}. 
We adopt the \citet{2011schlafly}  value of $\rm{E(B-V) = 0.08}$ mag for the analysis in this paper and use the extinction law of \citet{1989cardelli} with $\rm{R_{V} = 3.1}$. 
\section{Light Curve Evolution} \label{sec:LCEvolve}
\begin{figure}
\begin{center}
\includegraphics[width=\columnwidth]{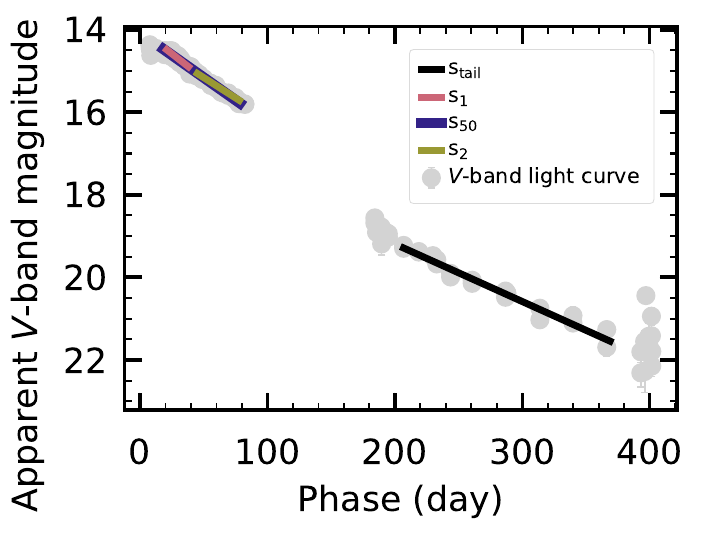} 
\caption{The $\rm{s_{1}}$ (pink), $\rm{s_{2}}$ (yellow), $\rm{s_{50V}}$ (indigo), and $\rm{ s_{tail}}$ (black) fits plotted over the period used to derive the relation. 
The {\it V}-band light curve is plotted in gray. 
Although errors in the apparent magnitude are plotted, in many cases these are smaller than the points.}
\label{fig:slopes}
\end{center}
\end{figure}

The light curve of ASASSN-15oz reaches peak brightness in the {\it V}-band 8.25 days after explosion, one day after the start of our observations, reaching an above average absolute magnitude of -18.05 $\rm{\pm}$ 0.025 mag.
The early light curve rises until around day 20 in the redder filters, flattening in the {\it V}-band and falling in the blue bands during this same period. 
It then falls in all bands for the remainder of the observations, changing slope around day 40 and again sometime after day 85, as it enters the radioactive decay phase, following the fall from plateau. 
Following \citet{2014anderson}, we use the {\it V}-band light curve to fit a slope to the first steep initial decline after maximum ($\rm{s_1}$) and to the second shallower slope prior to the fall from plateau ($\rm{s_2}$). 
We also fit a global decline rate, $\rm{s_{50v}}$ from soon after maximum to prior to the fall from plateau.
During the nebular phase we fit a slope to the radioactive decay tail ($\rm{s_{tail}}$\footnote{this is equivalent to $\rm{s_{3}}$ in \citet{2014anderson}}). 
The slopes we fit as well as the details of the fit are summarized in Table \ref{tab:slope} and plotted on top of the {\it V}-band light curve in Figure \ref{fig:slopes}.
Both the high slopes and bright maximum luminosity of ASASSN15oz are similar to those of other IIL-like SNe. 
\begin{table*}
\centering
\caption{The best-fit slope to the V-band light curve of ASASSN-15oz measured between the start and end phase listed in the table. All slopes are measured in units of magnitudes per 50 days.\label{tab:slope}}
\begin{tabular}{ccccc}
\hline
Slope Type & Slope (mag $\rm{(50 days)^{-1}}$) & Slope Error (mag $\rm{(50 days)^{-1}}$) & Start Phase (day) & End Phase (day) \\
\hline
$s_1$ & 1.21 & 0.066 & 21.0 & 38.0 \\
$s_2$ & 0.99 & 0.021 & 45.0 & 78.0 \\
$s_{50V}$ & 1.09 & 0.014 & 18.4 & 78.0 \\
$s_{tail}$ & 0.70 & 0.029 & 206.9 & 368.0 \\
\hline
\end{tabular}
\end{table*}
The slope of the radioactive decay tail is steeper than expected from the radioactive decay of $\rm{{}^{56}Ni \rightarrow {}^{56}Co \rightarrow {}^{56}Fe}$ implying that the ejecta is not fully trapping and reprocessing the gamma rays produced by the radioactive decay.
\\
\indent The pseudo-bolometric luminosity is computed following \citet{2008valenti}.
To summarize the method, the apparent magnitudes are corrected for galactic extinction.
We use  the {\it U, B, g, V, r, R, i,} and {\it I}-bands to calculate the psuedo-bolometric luminosity at the {\it V}-band cadence.
The flux density is calculated from the apparent magnitude at the effective wavelength of each filter and integrated using Simpson's Rule. 
Finally, the flux is converted to a luminosity using the distance modulus. 
When {\it V}-band observations extended either earlier or later than a filter, we use the color with the neighboring filter to extrapolate the flux.
\subsection{Nickel mass determination}\label{Sec:Ni}
The amount of  $\rm{{}^{56}Ni}$ synthesized during the explosion can be inferred using the late time ($\rm{>}$ 150 days) photometry.
At this time, the energy output is powered by the radioactive decay of $\rm{{}^{56}Ni \rightarrow {}^{56}Co \rightarrow {}^{56}Fe}$ which produces gamma rays that are reprocessed by the ejecta producing optical light.
Assuming that the SN has the same SED as the extremely well characterized SN 1987A, the $\rm{{}^{56}Ni}$ mass synthesized in the SN explosion can be found by scaling the pseudo-bolometric luminosity of SN when it is powered by $\rm{{}^{56}Co}$ to that of SN 1987A with the following equation \citep{2014spiro}:
\begin{equation} \label{eqn:Ni}
M({}^{56}Ni) = 0.075M_{\odot} \times \frac{L_{SN}(t)}{L_{87A}(t)}
\end{equation}
where $\rm{M(\,{}^{56}Ni)}$ is the synthesized $\rm{{}^{56}Ni}$ mass, $\rm{L_{SN}(t)}$ is the pseudo-bolometric luminosity of the SN at time t, and $\rm{L_{87A}(t)}$ is the pseudo-bolometric luminosity of SN 1987A at time t using the same filters as were used to compute the pseudo-bolometric luminosity of the SN.

\begin{figure}
\begin{center}
\includegraphics[width=\columnwidth]{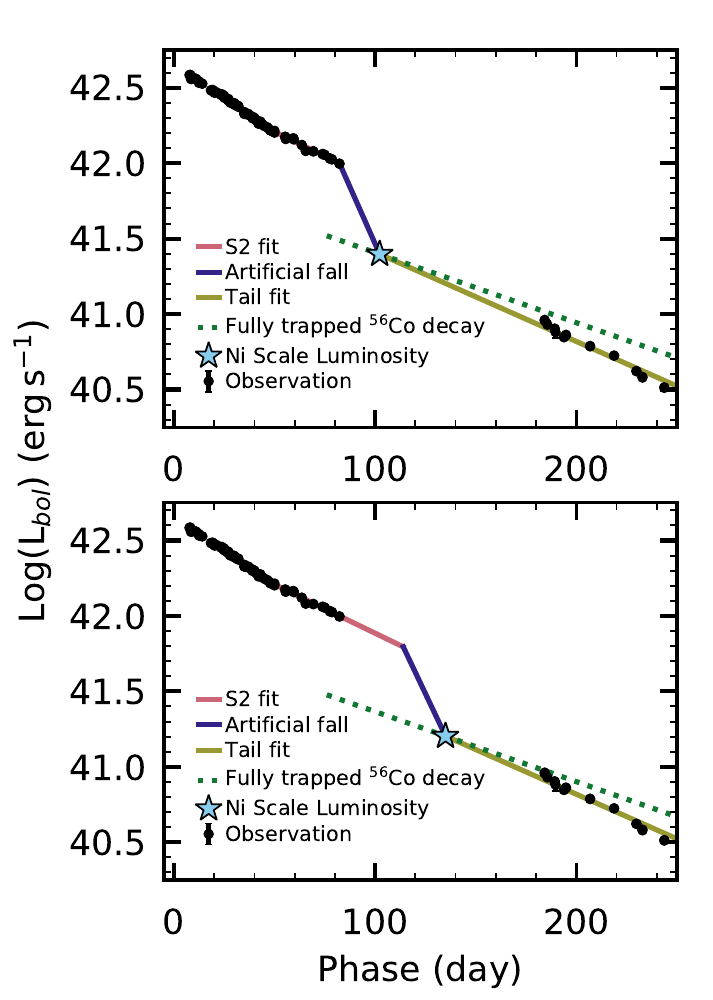} 
\caption{The artificial light curve of ASASSN-15oz for the shortest (top) and longest (bottom) plateau lengths. 
The black points plot the bolometric luminosity while the pink line represents a fit to the s2 data, the indigo line is the artificial fall from plateau, and the yellow line is a fit to the radioactive decay tail.
The cyan star marks the point to which the light curve of SN 1987A is scaled to derive a $\rm{{}^{56}Ni}$ mass. 
Plotted as a green dashed line is the slope of the radioactive decay tail if there was complete trapping.
An upper limit on the $\rm{{}^{56}Ni}$ mass of 0.11 \msun is derived from the light curve in the top panel while a lower limit of 0.08 \msun is derived from the lower panel. }
\label{fig:Ni}
\end{center}
\end{figure}

This scaling can be performed at any time, as long as the slope is the same as that of SN1987A, i.e. consistent with the complete trapping of gamma rays.
As the SN ages, the optical depth of the gamma rays decreases.
Thus a SN is most likely to have complete trapping immediately after the fall from plateau. 
Unfortunately, for ASASSN-15oz, there is not complete trapping for any part of the observed radioactive decay tail and there are no observations of the period immediately after the fall from plateau. 
We therefore consider the maximum and minimum $\rm{{}^{56}Ni}$ mass allowed by the observations, under the assumption that there is complete trapping immediately following the fall from plateau. 
For both cases we also assume that the fall from plateau takes 20 days, based on light curve fits presented in \citet{2016valenti}. 
An upper limit on the $\rm{{}^{56}Ni}$ mass is obtained by assuming ASASSN-15oz fell from plateau immediately after the last photospheric point  (see upper panel of Figure \ref{fig:Ni}).
Scaling the psuedo-bolometric luminosity of SN1987A at 102 days to that of ASASSN-15oz at the same time, we calculate a $\rm{{}^{56}Ni}$ mass of 0.11 \msunperiod.
To find the lower limit, we estimate a conservative value of the time to the middle of the fall from plateau ($\rm{t_{pt}}$) to be 125 days using Figure 5a in \citet{2016valenti}.
Using this $\rm{t_{pt}}$ we calculate the fall from plateau to end 135 days after explosion and extrapolate the tail fit to this phase (see lower panel of Figure \ref{fig:Ni}). 
We compute a $\rm{{}^{56}Ni}$ mass of 0.08 \msun by scaling the psuedo-bolometric luminosities at 135 days.
In both panels of Figure \ref{fig:Ni}, the observations are represented with black circles, the $\rm{s_{2}}$ slope with a pink line, the fall from plateau with a indigo line, the tail slope with a yellow line, and the expected tail slope from complete trapping is marked with a dashed green line. 
The time that is used to scale SN 1987A to ASASSN-15oz is marked with a cyan star.
\section{Spectroscopic Analysis} \label{sec:SpecEvolve}
The spectroscopic evolution of a SN during the photospheric phase provides insight into details of the SN ejecta.
The changes in velocity and the individual line profiles describe the geometry and energetics of the ejecta. 
The presence of different species at different times gives information about the chemical composition, temperature, and density of the ejecta.
In this section we analyze optical and NIR spectra.
\subsection{Optical Evolution} \label{OpticalEvolve}
\begin{figure*}
\begin{center}
\includegraphics[height=\textwidth]{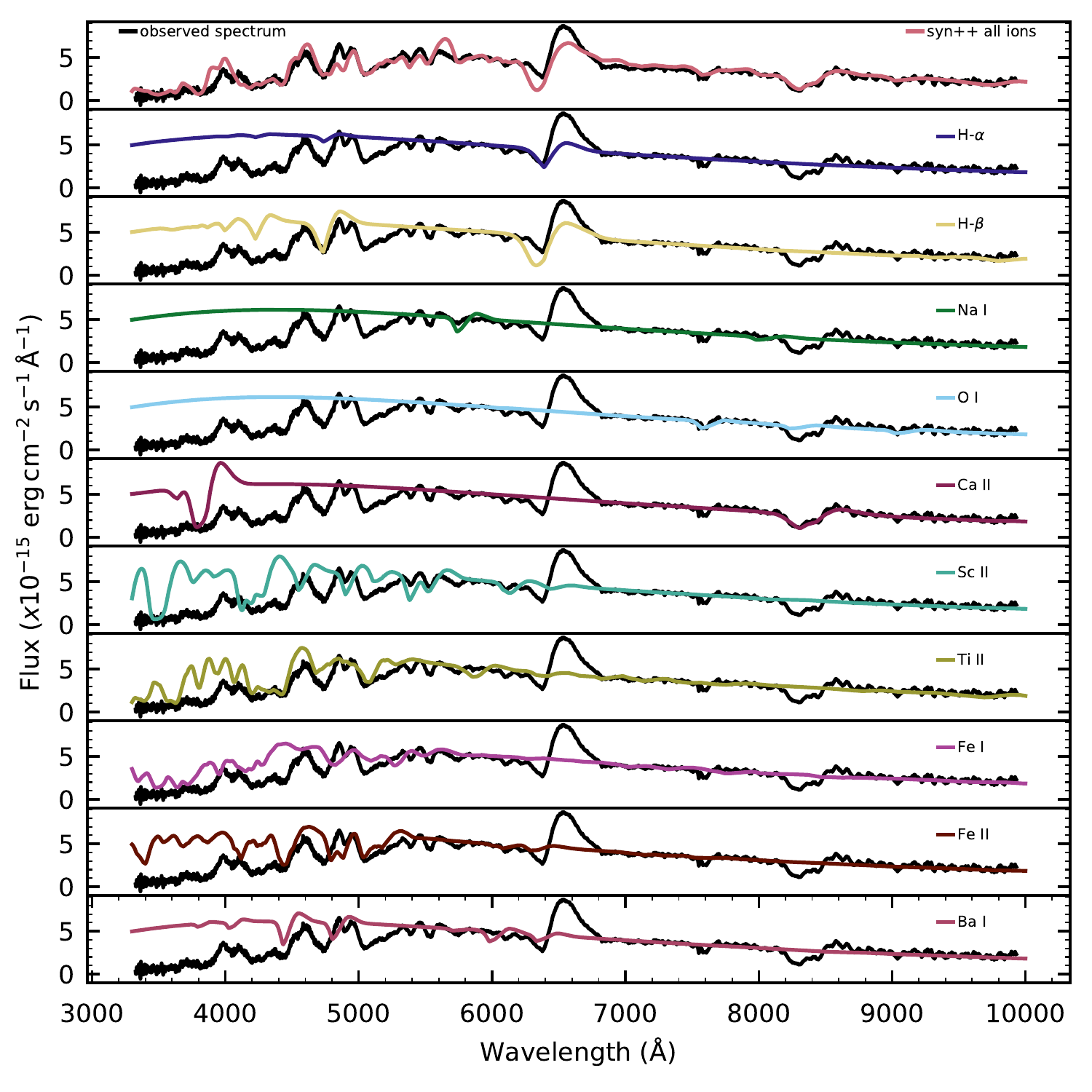} 
\caption{A comparison of the model spectra produced by Syn++ (colored lines) and the observed spectrum (black) from day 40 (2015 Oct 06).
The top panel shows the best fit with all elements while the remaining panels show the fit for each element individually.
We fit H $\rm{\alpha}$ and H $\rm{\beta}$ separately as H $\rm{\alpha}$ is not well modeled by the pure resonance scattering assumed in Syn++.
The H $\rm{\beta}$ fit is used in the combined spectrum in the top panel.}
\label{fig:syn++}
\end{center}
\end{figure*}

Initial line identification was performed by comparing the spectra to that of SN 1999em \citep{2001leonard}. 
Identification was confirmed using the spectrum synthesis code SYN++ \footnote{\citet{2013thomas}}(\citealt{2011thomas}; parameter details in Table \ref{tab:syn++}). 
The results of the SYN++ fit can be seen in Figure \ref{fig:syn++}.

We use two methods to find the velocity of lines for which individual components can be resolved, one for blended features, and one for individual features.
For lines that are not blended or that are dominated by the absorption of one ion (e.g. Sc II ($\rm{\lambda}$ 5526, 5662, 6262)), we employ the method presented in \citet{2012silverman}. 
For blended features we fit a multi-component gaussian to the feature.
We use this method to fit H $\rm{\alpha}$ ($\rm{\lambda}$ 6561), O I ($\rm{\lambda}$ 7774), FeII ($\rm{\lambda}$ 5169), and Ca II ($\rm{\lambda\lambda\lambda}$ 8498, 8542, 8662). 
Details of both fitting procedures are given in Appendix \ref{AppLineFit}.
Rather than expressing an uncertainty on the fit, we calculate the range of velocities that encompass 68.2\% of the integrated (continuum subtracted) flux around the feature minimum.
Because each line is formed in an extended region of the ejecta, this represents the range of velocities at which each line forms.

\begin{figure}
\begin{center}
\includegraphics[width=\columnwidth]{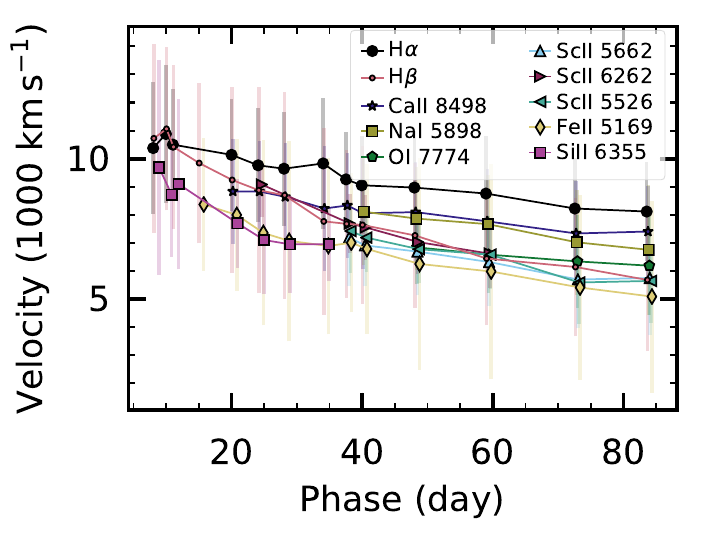}
\caption{The evolution of the velocity of H $\rm{\alpha}$, H $\rm{\beta}$, Na I, O I, Ca II, Sc II, and Fe II over time. 
The points represents the minimum of the feature showing the velocity with the largest optical depth while the bars plot the range of velocities found in the line forming region.}
The coincidence of velocities of different species, especially at late times, indicates that mixing may have been important for ASASSN-15oz.
\label{fig:velocity}
\end{center}
\end{figure}

\begin{figure*}
\begin{center}
\includegraphics[width=\textwidth]{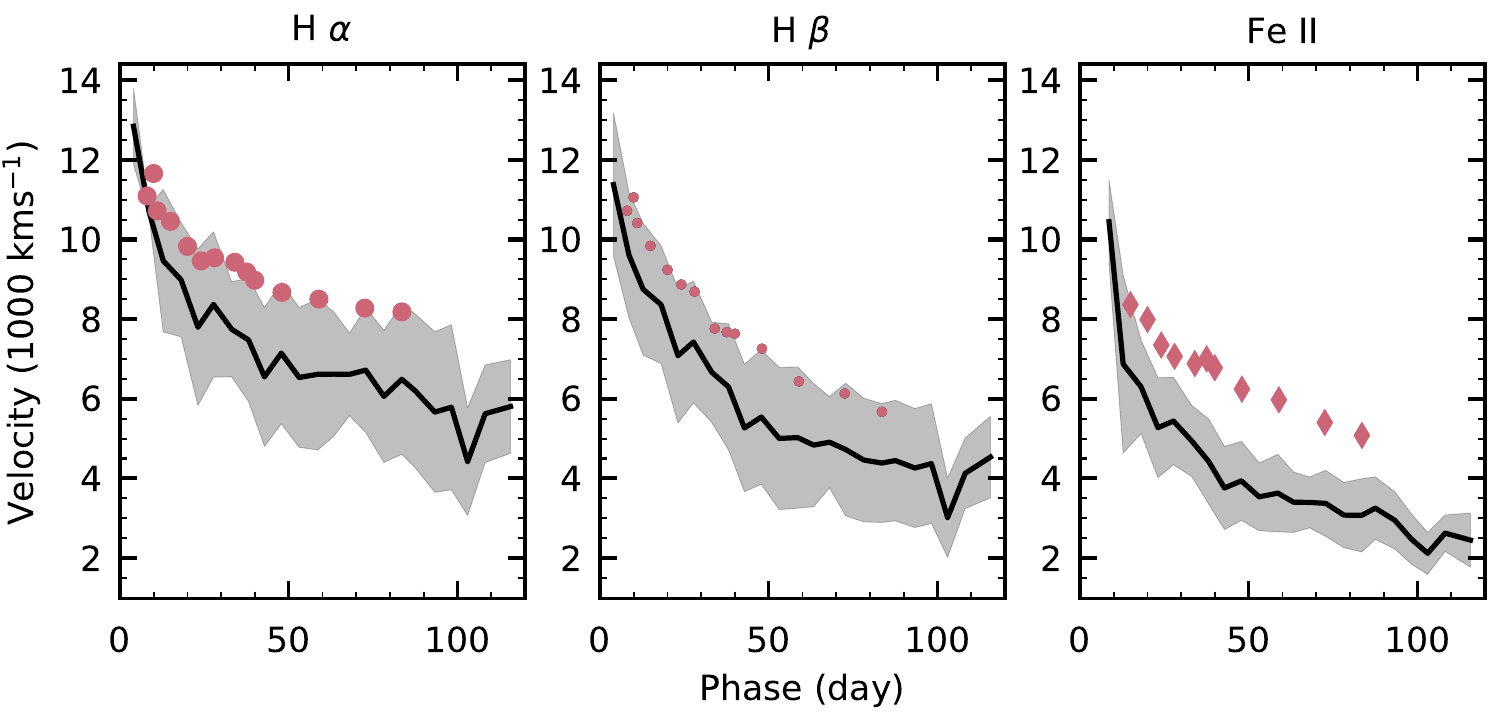}
\caption{A comparison of the velocity of ASASSN-15oz (pink) to the mean velocity of 122 IIP/IIL SNe \citep{2017gutierrez} (black) for H $\rm{\alpha}$ (left), H $\rm{\beta}$ (middle), and Fe II ($\rm{\lambda 5169}$; right). 
The gray regions represent the standard deviations of the mean velocities.
The velocity of ASASSN-15oz is $\rm{>1\sigma}$ above average for all three ions. 
Each of these line originates in a different part of the ejecta indicating that this is a global trend and the explosion energy is above average. 
Following \citep{2012silverman} we select velocity errors of 2 \AA.
These are contained within the symbols and so are not plotted.}
\label{fig:VelocityCompare}
\end{center}
\end{figure*}

Figure \ref{fig:velocity} shows the measured velocities. 
The distribution of velocities is typical of other IIP/IIL SNe representing the distribution of elements throughout the ejecta. 
A comparison of the velocities derived in this paper to the average values of 122 IIP/IIL SNe described in \citet{2017gutierrez} is shown in Figure \ref{fig:VelocityCompare}. 
The shaded regions represent the standard deviation of the sample.
We find above average velocities throughout the ejecta.
\subsection{Infrared Evolution}
Line identification was performed by comparing the NIR SOFI and SpeX spectra to those in \citet{2015valenti} and \citet{2018tomasella}.
Figure \ref{fig:IRmontage} shows the NIR spectra with prominent features labeled.
Like the optical, the first spectrum, taken eight days post explosion (2018 Sep 05), is dominated by hydrogen features.
The later spectra also mirror the optical evolution, developing metal lines as the ejecta slows and cools.
We use the IR spectra to search for evidence of interaction via a high velocity He I feature (see Section \ref{sec:cachito}).
\subsection{Progenitor Mass from Nebular Spectra} \label{sec:nebular}
\begin{figure}
\begin{center}
\includegraphics[width=\columnwidth]{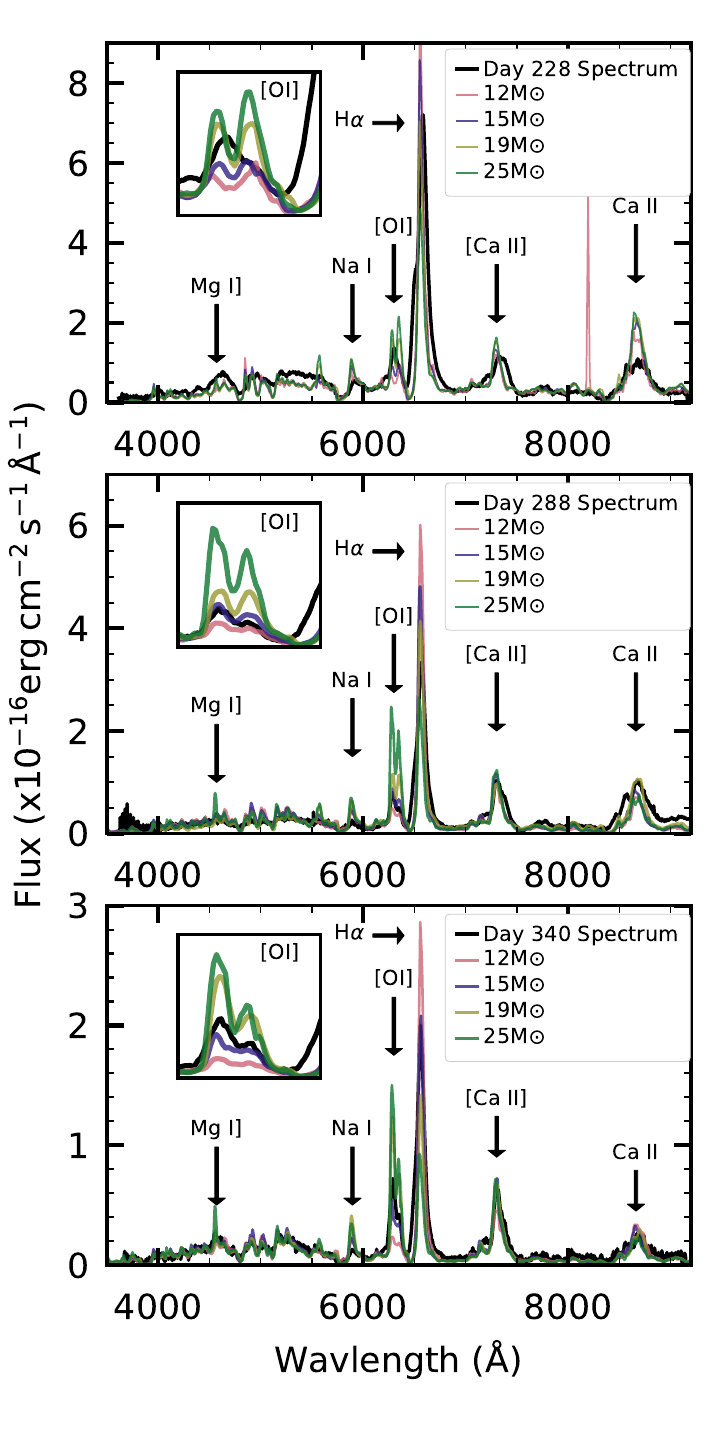} 
\caption{The nebular spectra of ASASSN-15oz from day 228 (2016 Apr 11; top), day 288 (2016 Jun 10; middle), and day 342 (2016 Aug 03; day 340; bottom) plotted with the scaled models of \citet{2014jerkstrand} for $\rm{M_{ZAMS}}$ = 12, 15, 19, and 25 M$\rm{_{\odot}}$. 
Most features in the observed spectra are well matched by the models. 
The inset shows the [O I] ($\rm{\lambda\lambda6300, 6334}$) doublet, the strength of which increases monotonically with progenitor mass. 
The [O I] flux falls between the 15 and 19 M$\rm{_{\odot}}$ models, consistent with the 18 M$\rm{_{\odot}}$ progenitor mass found using light curve modeling in Section \ref{sec:LCmodeling}.}
\label{fig:neb}
\end{center}
\end{figure}

Around 200 days the SN ejecta becomes optically transparent, revealing the inner core. 
At this time the SN enters the nebular phase, showing strong emission lines and no continuum (see Figure \ref{fig:neb}).
Analysis of the spectra at this time can constrain the geometry of the explosion as well as the abundance of different elements.
In particular, spectral modeling of this stage is a powerful tool to constrain the nature of the progenitor.
\citet{2012jerkstrand,2014jerkstrand} have shown that the intensities of a few lines are sensitive to the M$\rm{_{ZAMS}}$ of the progenitor.
Specifically, there is a tight monotonic correlation between the strength of the [OI] ($\rm{\lambda\lambda6300, 6334}$) line and the M$\rm{_{ZAMS}}$ of the progenitor.
This is because [OI] is more isolated and less sensitive to the explosive nucleosynthesis of the SN than the other lines and thus reflects the oxygen abundance of the progenitor which has been shown to correlate well with progenitor mass \citep{1995woosley}.

\citet{2014jerkstrand} modeled nebular spectra for 12, 15, 19, and 25 M$\rm{_{\odot}}$ progenitors ($\rm{M_{ZAMS}}$).
Starting with the explosion models from \citet{2007woosley}, the ejecta is divided into zones based on chemical composition and the evolution of the spectrum is found by modeling the gamma ray transport and deposition, non-thermal electron degradation, NLTE ionization and excitation, and Monte Carlo radiative transfer.

Four nebular spectra of ASASSN-15oz were taken between day 228 (2016 Apr 11) and  day 389 (2016-09-19).
The first three spectra are high S/N and are use for modeling.
In Figure \ref{fig:neb} we compare these epochs to the different mass models, scaling the models to the observations over the full wavelength range. 
A zoom in on the [OI] line is displayed in the inset of each panel. 
We find the [OI] strength of ASASSN-15oz falls between the 15 and 19 M$\rm{_\odot}$ models, consistent with the 17 M$\rm{_{\odot}}$ found by modeling the light curve (see Section \ref{sec:LCmodeling}).

As a sanity check of the comparison between observations and synthetic spectra, we infer the $\rm{{}^{56}Ni}$ mass from the scale factor we used to scale the models to the observations.
Following \citet{2018jerkstrand}, the scaling combined with the equation for the luminosity of the cobalt decay (Equation 6 of \citealt{2012jerkstrand}) yields the $\rm{{}^{56}Ni}$ mass (assuming complete trapping): 
\begin{equation}
\frac{F_{obs}}{F_{mod}} = \frac{d_{mod}^{2}}{d_{obs}^{2}}\frac{(M_{56,Ni}){}_{obs}}{(M_{56, Ni}){}_{mod}}exp\left(\frac{t_{mod} - t_{obs}}{111.4}\right)
\end{equation}
where $\rm{F_{obs}}$ and $\rm{F_{mod}}$ are the observed and model fluxes, $\rm{d_{obs}}$ and $\rm{d_{mod}}$ the observed and model distances, $\rm{(M_{56, Ni}){}_{obs}}$ and $\rm{(M_{56, Ni}){}_{mod}}$ are the observed and model $\rm{{}^{56}Ni}$ masses, and $\rm{t_{obs}}$ and $\rm{t_{mod}}$ are phases of the observation and the model.
Using this relation, we average the $\rm{{}^{56}Ni}$ mass found for each mass model to find an $\rm{{}^{56}Ni}$ mass of 0.03 \msun for day 228 (2016 Apr 11) and day 288 (2016 Jun 10) and 0.02 \msun for day 340 (2016 Aug 03).
These values are a factor of four smaller than our lower limit for the $\rm{{}^{56}Ni}$ mass (see Section \ref{Sec:Ni}).
This is due at least in part to the fact that the spectral synthesis modeling does not account for the incomplete gamma ray trapping in ASASSN-15oz.
The fact that several IIP/IIL SNe do not have complete trapping (see Section \ref{SecComp}) should be taken into account in the future.
\section{Evidence for interaction} \label{sec:Interaction}
Although it is well known that massive stars undergo significant mass-loss during their lives, the details of this mass-loss (timing, rate, speed, geometry) are not well understood.
The location and density of circumstellar material around a star record the amount of mass lost at different times during a star's evolution. 
As the SN shock passes through this material, it provides a unique probe of this material and thus the mass-loss of the progenitor.
While the mass-loss of stars with dense CSM (e.g. IIn SNe) that show strong evidence of interaction, such as narrow emission lines in photospheric spectra or irregular photometric brightening, is often studied, the mass-loss of SNe with lower density CSM is often ignored.

In IIP-like SNe, the plateau is the result of the recession of the photosphere due to hydrogen recombination matching the expansion of the SN ejecta.
The linear decline of the light curves of IIL-like SNe is then explained by the photosphere receding faster than the ejecta's expansion.
The photosphere's recession is governed by the rate at which the ejecta becomes optically thin.
This rate is determined by either the expansion velocity of the ejecta or the amount of hydrogen in the ejecta, or both. 
If the expansion velocity is above average, the ejecta becomes optically thin faster, leading to a more IIL-like light curve.
Likewise, if there is less hydrogen in the ejecta, then the recombination front moves more rapidly through the ejecta, leading to a more IIL-like light curve.
IIL-like SNe show above average ejecta velocities and weaker than average hydrogen absorption \citep{2014gutierrez}, implying that both of these effects are at play.
Therefore, the linear shapes of the light curves of IIL-like SNe indicate that their progenitors have smaller hydrogen envelopes.
This then implies that IIL-like SNe should undergo more mass-loss than IIP-like SNe.
ASASSN-15oz, a IIL-like SN with photometric and spectroscopic observations from the X-ray through the radio is the ideal SN on which to study the importance of mass-loss for IIP-like and IIL-like SNe.
The different epochs and wavelengths probe different periods of mass-loss allowing for the characterization of the mass-loss history of the progenitor.
In this section we analyze the evidence for interaction at different wavelengths.
We note that we cannot search for flash ionization features as there are no spectroscopic observations until near maximum.
\subsection{UV and X-Ray}
ASASSN-15oz is not detected in any of the \textit{Swift} X-ray observations. 
To determine an upper limit on the X-ray luminosity, we sum all observations and find a limiting count rate of  $\rm{5.775\times10^{-4}}$ counts $\rm{s^{-1]}}$ (0.4-5 keV).
Assuming Galactic absorption of 6.18x$\rm{10^{20}}$ cm$\rm{^{-2}}$ \citep{2005kalberla}, this corresponds to an unabsorbed flux of $\rm{1.647\times10^{-14}}$ erg cm$\rm{^{-2}}$ $\rm{s^{-1}}$ and adopting a distance of 28.83 Mpc, a luminosity upper limit of $\rm{2.56 \times10^{39}}$erg $\rm{s^{-1}}$.

We compare this upper limit to the model of SN 1999em \citep{2007chugai} which has similar progenitor and explosion parameters. 
They model the interaction as an infinitely thin double-shock structure \citep{1982chevalier, 1985nadyozhin} composed of a forward shock propagating into the surrounding gas and reverse shock propagating backwards into the SN ejecta, towards its core. 
\citet{2007chugai} find that the reverse shock dominates the X-ray observations for a pre-explosion mass loss rate of $\rm{10^{-6}}$ \msunperiod $\rm{yr^{-1}}$ and a wind velocity of $\rm{10}$ km $\rm{s^{-1}}$ producing luminosities between $\rm{10^{38}}$ and $\rm{10^{39}}$ erg $\rm{s^{-1}}$. 
The forward shock produces luminosities comparable to the reverse shock in the first 10 days and decreasing to below $\rm{10^{34}}$ erg $\rm{s^{-1}}$ by day 100.  
Our non-detection is at the upper limit of this range, making it plausible that if there is X-ray emission, it is below our detection limits.

We use SYN++ to model and tentatively identify absorption features from Fe II, Ti II, and Mg II features in the first {\it Swift} UV grism spectrum (see Figure \ref{fig:SwiftSpectrum}). 
This spectrum shows no sign of the narrow emission lines typically associated with strong interaction ( $\rm{\dot{M}>10^{-4}}$ g $\rm{cm^{-1}}$; \citealt{2012kiewe})
\subsection{Optical Spectroscopy} \label{sec:cachito}
\citet{2007chugai} proposed a high velocity hydrogen and helium line are produced by interaction of the SN ejecta with the stellar wind. 
According to their model, a notch on the blue side of the H $\rm{\alpha}$ profile should appear 40-80 days post explosion.
Although they could construct a model with the introduction of a cold dense shell in which a similar notch was visible in H $\rm{\beta}$ their simplest model did not show H $\rm{\beta}$ due to its low opacity in the wind.
They do however, predict absorption in He I $\rm{\lambda10830}$ 20-60 days post explosion, a line that is only excited when wind is present. 

\begin{figure}
\begin{center}
\includegraphics[width=\columnwidth]{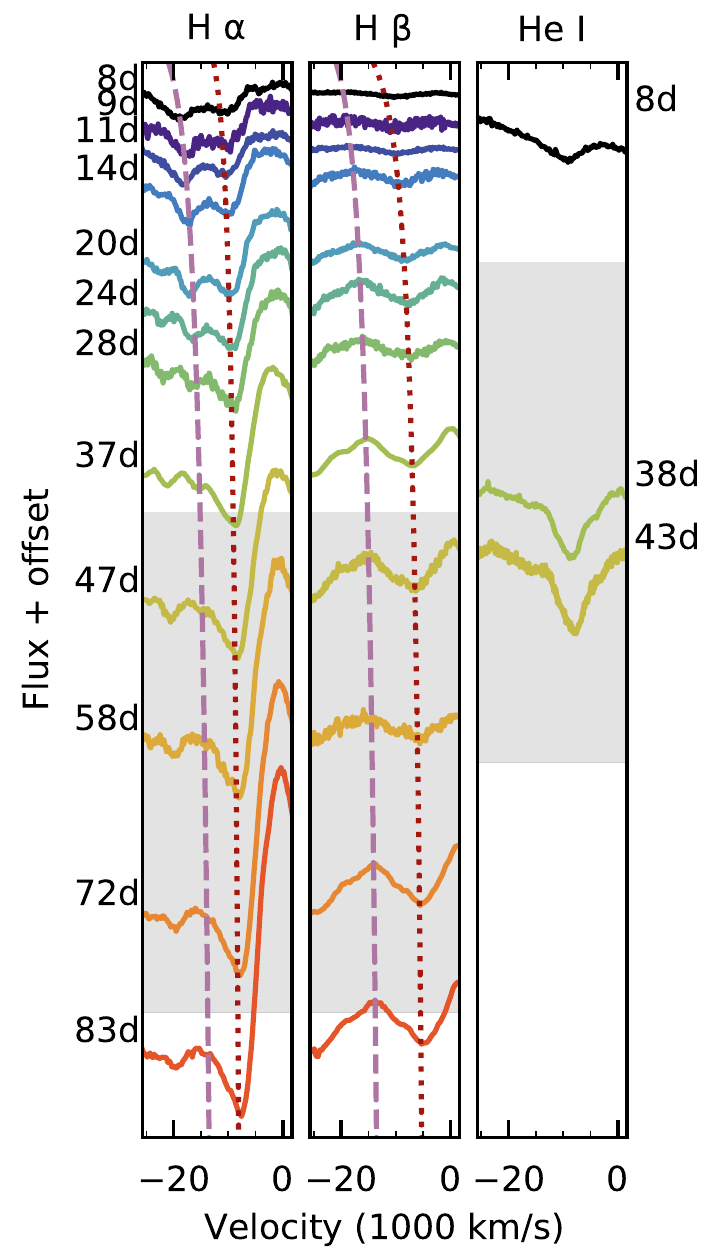}
\caption{The evolution of the cachito feature (left panel; dashed line) over time. 
The feature is visible bluewards of H $\rm{\alpha}$ (left panel; dotted line)  in the first spectrum, eight days post explosion.
It further increase in strength for the next 10 days, then decreases until it is barely visible at 60 days post explosion.
This evolution is counter to the evolution described in \citet{2007chugai} in which the high velocity hydrogen feature becomes visible around day 40, making it unlikely that it is high velocity hydrogen.
There is no evidence of a high velocity hydrogen feature in H $\rm{\beta}$ (center panel; H $\rm{\beta}$ marked with dotted line; cachito velocity marked with dashed line) although this is not surprising given the low opacities predicted by \citet{2007chugai}.
The high velocity He I ($\rm{\lambda=10830}$) feature (right panel) also predicted by the models of \citet{2007chugai} is too heavily blended with the Paschen-$\rm{\gamma}$ and C I lines to be identifiable.}
\label{fig:CachitoEvolve}
\end{center}
\end{figure}

\begin{figure}
\begin{center}
\includegraphics[width=\columnwidth]{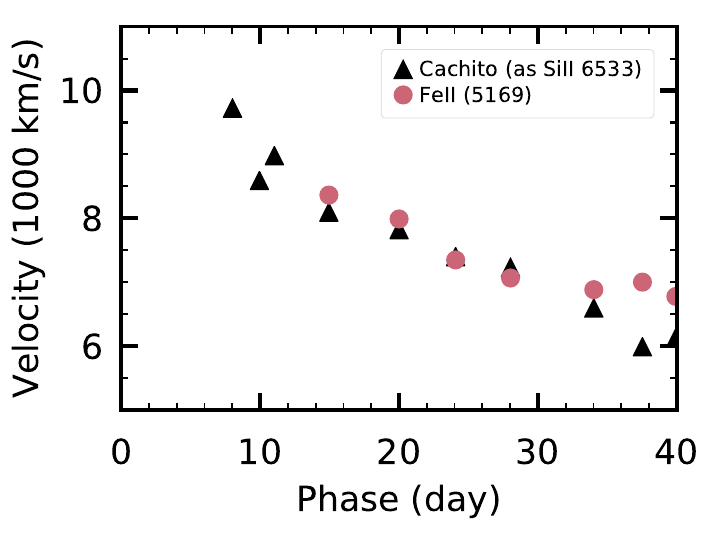}
\caption{A comparison of the velocity of the cachito feature if it is Si II ($\rm{\lambda6355}$) (black triangles) and the velocity of the metals in the ejecta as characterized by the Fe II $\rm{\lambda5169}$ lines (pink circles).
The velocities are the same indicating that this line is likely due to Si II rather than high velocity hydrogen in the circumstellar medium.}
\label{fig:SiVelocity}
\end{center}
\end{figure}

A feature blueward of H $\rm{\alpha}$ is often present in Type II SNe, however, it is unclear if this feature is due to high velocity hydrogen or to Si II ($\rm{\lambda}$6355)
and it has thus been named the cachito feature to avoid association with a particular species \citep{2017gutierrez}.
Building on a the analysis of smaller samples (e.g. \citealt{2001leonard}, \citet{2013inserra}, \citealt{2014valenti}), \citet{2017gutierrez} searched for this feature in 122 IIP/IIL SNe. 
They find the cachito feature in 70 SNe in their sample and divide the detections in early phase detections (phase < 40 days) and late phase detections (phase > 40 days).
The velocities of the cachito feature observed in the early sample are most often well matched to the Fe II velocity if interpreted as Si II. 
For this reason, they interpret the cachito feature in the early sample as Si II and at late times as high velocity hydrogen.
Figure \ref{fig:CachitoEvolve} shows the time evolution of the region surrounding H $\rm{\alpha}$ (left panel), H $\rm{\beta}$ (middle panel), and He I (right panel).
H $\rm{\alpha}$ and H $\rm{\beta}$ are marked with a brick dotted line in the left and middle panels, respectively.
The dashed line in the left panel marks the location of the cachito feature and in the middle panel, the expected location of the cachito feature in H $\rm{\beta}$.
The time of expected CSM interaction is marked in grey.
ASASSN-15oz shows a strong cachito feature in the first spectrum, eight days post explosion. 
This feature strengthens until about day 15 and then fades until it is no longer visible at day 60. 
We find no evidence for a high velocity feature in H $\rm{\beta}$.
The He I feature is fully blended with C I ($\rm{\lambda 10691}$) and Paschen-$\rm{\gamma}$ and if it is present, it is impossible to deblend and identify.
A comparison of the cachito velocity, when interpreted as Si II, and the Fe II velocity is plotted in Figure \ref{fig:SiVelocity}.
We find excellent agreement between the two velocities strengthening our conclusion that this feature is due to Si II rather than high velocity hydrogen.

\begin{figure}
\begin{center}
\includegraphics[width=\columnwidth]{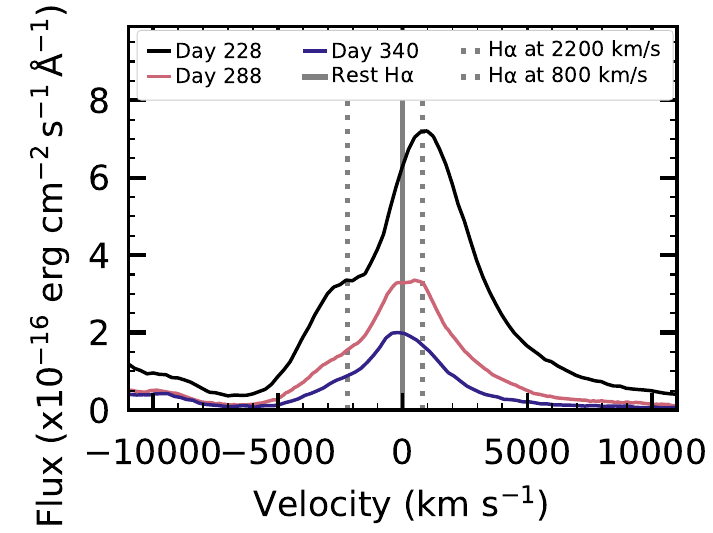} 
\caption{The asymmetric H $\rm{\alpha}$ emission in the nebular spectra of ASASSN-15oz. 
The spectrum shows H $\rm{\alpha}$ emission at rest (solid gray line), blueshifted by 2200 km $\rm{s^{-1}}$ and redshifted by 800 km $\rm{s^{-1}}$ (dashed gray lines).
Although the asymmetry is visible in all of the spectra, it is most prominent in the earliest nebular spectrum on day 228 (2016 Apr 11) and could be a sign of interaction with CSM.
}
\label{fig:nebHa}
\end{center}
\end{figure}

The H $\rm{\alpha}$ emission profiles in the nebular spectra of ASASSN-15oz are asymmetric (see Figure \ref{fig:nebHa}).
In addition to emission at the rest wavelength of H $\rm{\alpha}$, there is evidence of both blueshifted and redshifted emission at 2200 km$\rm{s^{-1}}$ and 800 km$\rm{s^{-1}}$, respectively. 
This feature is strongest in the first spectrum at day 228 (2016 Apr 11), becoming weaker in the subsequent spectra.
Asymmetric H $\rm{\alpha}$ emission has been seen in a few IIP/IIL SNe (e.g. 1999em: \citealt{2001leonard}; SN 2004dj: \citealt{2005chugai}; SN 2013ej: \citealt{2017utrobin}).
\citet{2001leonard} explain the shape of the H $\rm{\alpha}$ emission as either due to an asymmetric line-emitting region or due to interaction between the ejecta and the CSM while \citet{2005chugai} and \citet{2017utrobin} attribute this feature to a bipolar $\rm{^{56}Ni}$ ejecta. 

\subsection{Radio}\label{sec:RadioInteraction}
\begin{figure}
\begin{center}
\includegraphics[width=\columnwidth]{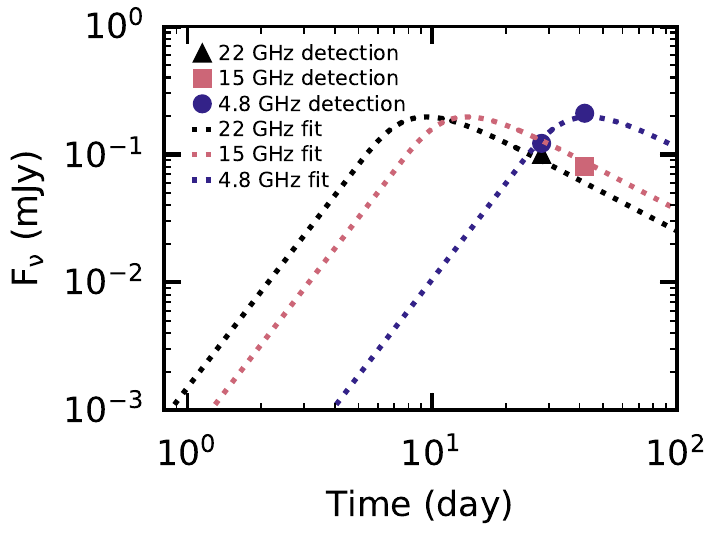}
\caption{The best fit SSA model (dotted lines) plotted over the observations (symbols) at 22 GHz (black), 15 GHz (pink), and 4.8 GHz (indigo).
These data are best fit by a shock velocity of $\rm{\sim 1.4\times10^{4}}$ km $\rm{s^{-1}}$ and a density of $\rm{K = 4.51\times 10^{11}}$ g $\rm{cm^{-1}}$}
\label{fig:radio}
\end{center}
\end{figure}

The origin of radio emission from SNe can be explained by the interaction of the SN ejecta with the CSM. 
This interaction leads to a shockwave traveling via the CSM that in turn accelerates electrons and enhances magnetic fields and thus synchrotron emission ensues \citep{1982chevalier,1998chevalier, 2002weiler, 2006chevalier}. 
It has been shown over the last decade that in most observed SNe, the optically thick part of the observed radio spectrum can be modeled by a synchrotron self absorption (SSA; e.g. \citealt{2012soderberg, 2013horesh}).
We use the \citet{1998chevalier} SSA formalism to model the radio measurements of ASASSN-15oz.
In such a model the radio emission peaks at a frequency below which the emission is strongly absorbed by SSA. 
Above that frequency the emission becomes optically thin. 
As the shockwave progress outwards to lower CSM densities, the SSA optical depth drops and the peak of the radio emission moves to lower and lower frequencies. 
We use Equation 1 from \citet{1998chevalier} to model the radio dataset (see Figure \ref{fig:radio}).
We use the standard equipartition assumption and set the microphysical parameters, $\rm{\epsilon_{e} = \epsilon_{B}  = 0.1}$ \citep{2006chevalier}. 
We also set the electron power-law distribution to $\rm{p=3}$, leaving the CSM density and shockwave velocity as free parameters.
We find the shockwave velocity (assuming constant velocity) is $\rm{\sim 1.4\times10^{4}}$ km $\rm{s^{-1}}$ (which is the typical average value observed in SNe; e.g. \citealt{2006chevalier}).
Assuming CSM surrounding the SN is created by a stellar wind from the progenitor prior to explosion, we model the density profile of the CSM as:
\begin{equation} \label{eqn:density}
\rho(r) = \frac{\dot{M}}{4\pi r^{2}v_{wind}} = \frac{K}{r^{2}}
\end{equation}
where $\rm{\dot{M}}$ is the wind mass-loss rate and $\rm{v_{wind}}$ is the wind velocity.
We find $\rm{K = 4.51\times 10^{11}}$ g $\rm{cm^{-1}}$.

With a wind velocity between 10-100 km $\rm{s^{-1}}$, this implies a mass-loss rate between $\rm{\dot{M}}$ $\rm{\approx 0.9-9\times 10^{-7}}$ \msunperiod $\rm{yr^{-1}}$.
\subsection{Light Curve Modeling with SNEC}\label{sec:LCmodeling}
We fit the light curve of ASASSN-15oz using the Supernova Explosion Code (SNEC) \citep{2015morozova}.
SNEC is an open source Lagrangian 1D radiation hydrodynamic code that employs flux-limited radiation diffusion and assumes local thermodynamic equilibrium (LTE).
SNEC uses a Paczy\'nski equation of state \citep{1983paczynski} solving for the ionization fractions using the Saha equations in the non-degenerate approximation \citep{2000zaghloul}. 
Opacities are drawn from OPAL type II opacity tables \citep{1996iglesias} at high temperatures ($\rm{T=10^{4.5}-10^{8.7}}$ K) and tables of \citet{2005ferguson} at low temperatures ($\rm{T = 10^{2.7}-10^{4.5}}$ K) supplemented with an opacity floor of 0.01 cm$\rm{^2}$g$\rm{^{-1}}$. 
The mixing of isotopes due to the explosion is modeled by boxcar smoothing the progenitor composition profile with a kernel of 0.4 \msunperiod.

In the past, hydrodynamic SN models that used progenitors evolved with stellar evolution codes (referred to as evolutionary progenitors for the remainder of this paper; e.g. {\it Kepler} code, \citealt{1978weaver,2007woosley,2015woosley, 2014sukhbold,2016sukhbold}; {\it MESA}, \citealt{2018paxton}) failed to reproduce the shape of the early observed light curves of Type II SNe. 
Some authors have solved this issue by using hydrostatic progenitor models with extended envelopes and density profiles that are set to match observations rather than calculated by stellar evolution codes (referred to as non-evolutionary progenitors for the remainder of this paper; e.g. \citealt{2008utrobin}, \citealt{2017utrobin}). 
However, adding a dense CSM ejected just before the explosion to progenitors evolved with stellar evolution codes has resolved the discrepancy between the early observed light curves and the light curves modeled using physically derived progenitors (e.g. \citealt{2018morozova,2015gezari,2018foerster}).
The presence of this CSM is corroborated by other observations of IIP/IIL SNe (e.g. \citealt{2017yaron,2018bullivant}).

For the progenitor models, we use a set of non-rotating solar metallicity RSGs parameterized by the $\rm{M_{ZAMS}}$, evolved with the {\it Kepler} code and described in \citet{2016sukhbold}.
Note that the stellar evolution calculations made with {\it Kepler} take into account the regular steady winds observed in RSGs, in the way prescribed by \citet{1990nieuwenhuijzen} and \citet{1999wellstein}.
For this reason, the final pre-SN masses of the models may be up to several $\rm{M_{\odot}}$ smaller than their initial $\rm{M_{ZAMS}}$.
Both the $\rm{M_{ZAMS}}$s and pre-SN masses of the models used are given in Table \ref{tb:param}.
However, for the typical mass loss rates of $\rm{\lesssim10^{-6}}$ \msunperiod $\rm{yr^{-1}}$,  the density in these regular steady winds is too low to have any noticeable effect on the post-explosion optical light curves, and these winds are not included in the progenitor profiles. 

We explore variations in CSM due to this enhanced mass-loss by adding a steady-state wind above the RSG models with the density profile expressed in Equation \ref{eqn:density}.
This approximation allows us to parameterize the CSM with two parameters, K and $\rm{R_{ext}}$, the radial extent of the CSM.
In reality, the late enhanced mass loss does not have to be in the form of a steady wind with constant $\rm{\dot{M}}$ and may instead represent one or several eruptive outbursts. However, in our experience, the early optical light curve depends weakly on the exact slope of the CSM profile, being more sensitive to the total CSM mass.

Recently, \citet{2018paxton} demonstrated that IIP/IIL SNe are well characterized by the radiation diffusion approximation from shock breakout through the fall from plateau.
At the same time, they have shown that the assumption of LTE is not well satisfied at the photosphere location of the IIP/IIL SNe models. 
For this reason, SNEC color light curves generally rise faster than the light curves obtained from more sophisticated multi-group radiation-hydrodynamics codes, like STELLA \citep{2018paxton}. 
However, this fact makes the case for introducing a CSM surrounding the RSG before its explosion even stronger.
In this view, the total CSM mass derived in our analysis may be considered as a lower limit, while the values obtained with more advanced codes should be comparable, or in some cases larger \citep{2017moriya,2018paxton}.

To account for the formation of the neutron star, we excise the inner $\rm{1.4\,M_{\odot}}$ of the progenitor models prior to the explosion. 
After that, we model the explosion of ASASSN-15oz using a thermal bomb with explosion energy $\rm{E_{exp}}$ lasting for one second in the inner 0.02 \msunperiod. 
When computing the amount of energy injected in the form of a thermal bomb, SNEC automatically takes into account the total initial (negative, mostly gravitational) energy of the models. 
Therefore, by the explosion energy $\rm{E_{exp}}$ in our analysis we mean the total energy of the models after explosion, which is conserved in the code to better than 1\% accuracy and eventually mainly transformed into the kinetic energy of the expanding envelope.

SNEC does not model nuclear reaction networks, but rather takes as input a mass of $\rm{{}^{56}}$Ni. 
We use a grid of $\rm{{}^{56}}$Ni masses informed by our analysis in Section \ref{15ozIntro}.
While SNEC allows for the $\rm{{}^{56}}$Ni to be mixed out to different values of mass coordinate, \citet{2017morozova} find that the progenitor masses and explosion energies derived from fitting the IIP/IIL SN light curves are not very sensitive to the degree of $\rm{{}^{56}}$Ni mixing.
Therefore, in our study we choose to keep this parameter fixed and mix the $\rm{{}^{56}}$Ni up to 5.0 M$\rm{_{\odot}}$.

We use SNEC to find the best progenitor parameters varying the progenitor mass, explosion energy, $\rm{{}^{56}}$Ni mass, CSM density, and CSM extent. 
Table \ref{tb:param} gives the set of parameters used, resulting in over 5000 model light curves.
\begin{table*}
\centering
\caption{The grid of parameters used by SNEC. 
The values that best fit the data are in bold. The pre-explosion masses are taken from Table 2 of \citet{2016sukhbold}.}
\label{tb:param}
\begin{tabular}{l|l}
\hline
Parameter & Values \\
\hline
Progenitor ZAMS Mass ($\rm{M_{ZAMS}}$; \msunperiod) & 11, 13, 14, 16, {\bf17}, 18, 21 \\
Final Pre-SN Mass ( \msunperiod) & 10.688, 11.567, 12.079, 13.145, {\bf14.301}, 14.936, 16.119 \\
Pre-SN Radius (100 R$\rm{_{\odot}}$)                     & 5.7, 7.0, 7.8, 8.9, {\bf9.1}, 9.7, 11.2 \\
Explosion Energy ($\rm{E_{exp}}$; 10$\rm{^{51}}$ ergs)   & 0.5, 0.8, 1.1, {\bf 1.4}, 1.7, 2.0 \\
CSM Density (K; 10$\rm{^{17}}$ g $\rm{cm^{-1}}$)            & 0, 10, 20, 30, 35, {\bf40}, 50, 60 \\
CSM Extent ($\rm{R_{ext}}$; 100 R$\rm{_{\odot}}$)           & 0, 15, {\bf18}, 21, 24, 27, 30, 33 \\
Ni Mass (\msunperiod)                                                        & {\bf 0.08}, 0.09, 0.11 \\
\hline
\end{tabular}
\end{table*}
\citet{2018morozova} find the best fit model with and without CSM as a two step process. 
First they modeling the second half of the light curve, characterized by the s2 slope without CSM to determine the best progenitor mass and explosion energy. 
Then, fixing the explosion energy and progenitor mass, they explore the CSM parameter space. 
This is computationally less intensive than modeling the full parameter space and allows them to explore a finer grid of parameters.
Given the complexity of the parameter space, we choose to model the best light curve with CSM, exploring the full set of parameters simultaneously. 

\begin{figure}
\begin{center}
\includegraphics[width=\columnwidth]{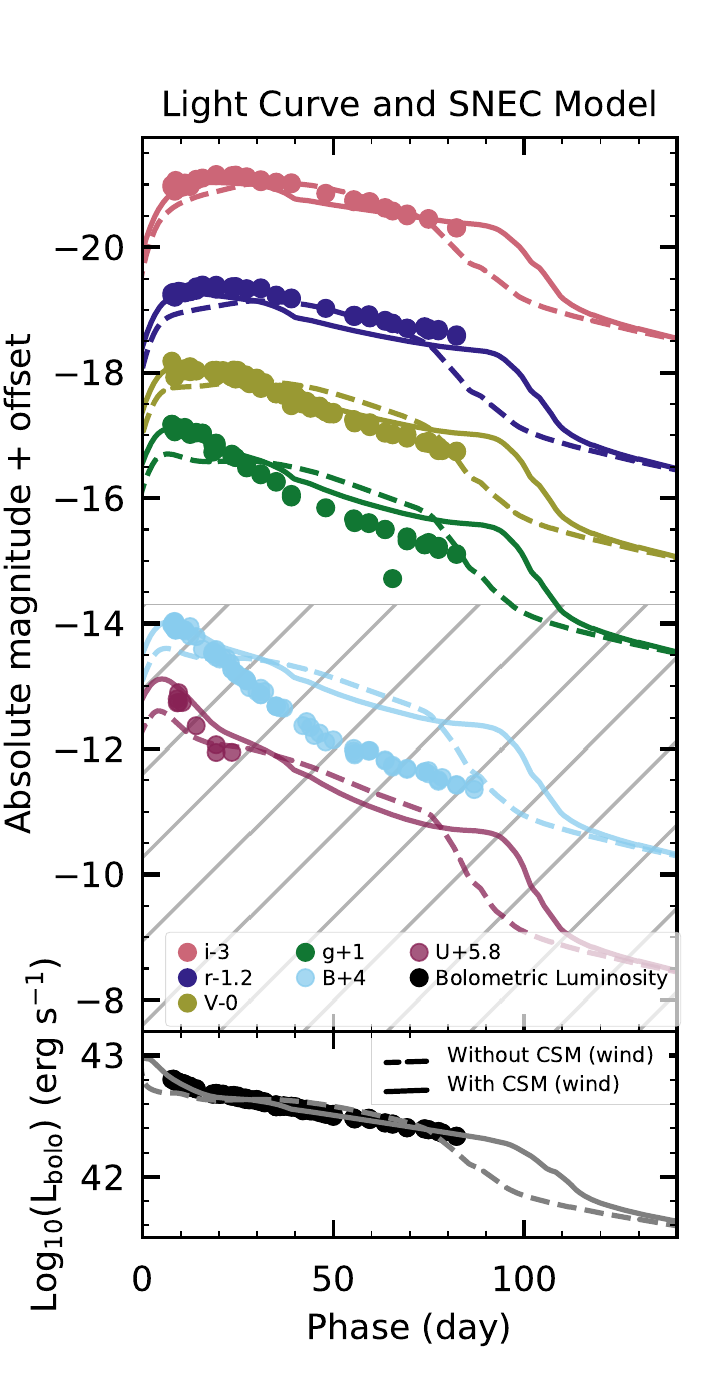} 
\caption{Top: the multi-band optical light curve of ASASSN-15oz. 
The observations are shown as circles and the best fit SNEC model is shown with and without CSM (solid and dashed lines, respectively).
We require CSM to fit the light curve at early times. 
The blue filters are affected by line blanketing ({\it U} and {\it V}-band) and are thus not expected to be well approximated by the blackbody approximation used to convert the bolometric luminosity produced by the modeling into magnitudes in individual filters. 
For this reason we do not use these in the fit and have placed hatched lines over them.
Line blanketing may also be responsible for the overestimate of the model flux in the {\it g}-band.
Bottom: the bolometric luminosity of ASASSN-15oz plotted with the best model fit to the bolometric luminosity including CSM (solid line) and the best model without CSM (dashed line).}
\label{fig:snecLC}
\end{center}
\end{figure}

For ease of comparison with observations, SNEC uses the photospheric temperature at each time step to compute a blackbody spectrum, which it combines with different filter throughputs to output a light curve in Sloan filters {\it u, g, r,} and {\it i}, Bessell filters {\it U, B, V, R,} and {\it I} and PanSTARRS filter {\it z}. 
During the rise and plateau phase, a blackbody should be a good approximation to the longer wavelengths. 
However, line blanketing may cause the bluer filters to be a poor representation of the observed spectrum \citep{2009kasen,2005dessart}.
For this reason the best fit model is determined using the {\it g, r,} and {\it i} filters. 
While we do have a {\it V}-band light curve, the throughput heavily overlaps with the {\it g}- and {\it r}-bands and its inclusion would give more weight to these wavelengths without providing new information.
The best fit model is determined by interpolating the well sampled model to the observed wavelengths and computing a chi-square minimization across all three filters.
Given the uncertainty in the explosion time, we shift the model spectrum by $\rm{\pm}$4 days and treat this offset as a free parameter (t$\rm{_{offset}}$).
Parameters for the best fit models with and without CSM are given in Table \ref{tab:BestModel}. 
These models are shown with (solid lines) and without (dashed lines) CSM interaction in the top panel of Figure \ref{fig:snecLC}. 
Integrating the CSM density over it radial extent, we find a total CSM mass of 1.5 \msunperiod. 
\begin{table*}
\centering
\caption{The best fit SNEC model parameters for the simultaneous fits to {\it g,r,} and {\it i}-bands and to the bolometric luminosity with and without CSM.}
\label{tab:BestModel}
\begin{tabular}{l|c|c|c|c|}
\hline
                                                                                            &\multicolumn{2}{|c|}{Fit to {\it g, r, i}-bands} &\multicolumn{2}{|c|}{Fit to Bolometric Luminosity}\\
\hline
Parameter                                                                          & With CSM            & Without CSM     & With CSM & Without CSM\\
\hline
Progenitor ZAMS Mass ($\rm{M_{ZAMS}}$; \msunperiod) &   17                    &     18                    &   18            &17\\
Final Pre-SN Mass ( \msunperiod)                          &   14.301             &      14.936            &     14.936    &14.301\\
Pre-SN Radius (100 R$\rm{_{\odot}}$)                    &  9.1                    &      9.7                  &       9.7       &9.1\\
Explosion Energy ($\rm{E_{exp}}$; 10$\rm{^{51}}$ ergs)    &    1.4                 &     2.0                 &     1.4             &2.0 \\
CSM Density (K; 10$\rm{^{17}}$ g $\rm{cm^{-1}}$)             &     40                  &     0                    &       60           &0 \\
CSM Extent ($\rm{R_{ext}}$; 100 R$\rm{_{\odot}}$)            &     1800              &     0                     &      1500       &0 \\
Ni Mass (\msunperiod)                                                         &     0.083               &    0.083              &    0.083       &0.083\\
Total CSM Mass (\msunperiod)                                            & 1.5                      &          0               &    1.4            &   0 \\
\hline
\end{tabular}
\end{table*}

As a sanity check, we derive the bolometric luminosity by fitting a black body to the photometric observations and find the best fit SNEC model with and without CSM. 
The parameters for these models are given in Table \ref{tab:BestModel} and are plotted over the bolometric luminosity in the bottom panel of Figure \ref{fig:snecLC}.
The explosion energy and ejecta masses are similar to those derived from a simultaneous fit of the {\it g, r,} and {\it i} bands. 
Although the individual CSM parameters vary, these are highly degenerate and the total CSM mass, which our model is a better measure of, is very similar.
The difference in the total CSM mass from the two different methods can be seen as an indication of the uncertainty in this value.

\begin{figure}
\begin{center}
\includegraphics[width=\columnwidth]{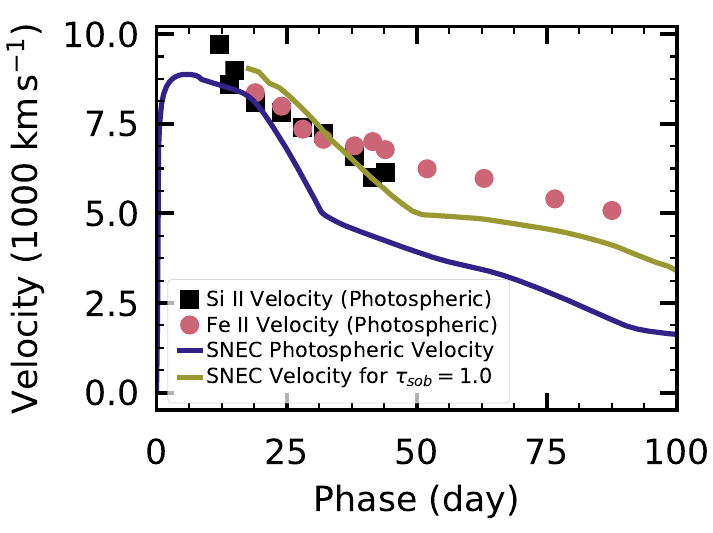} 
\caption{The model photospheric velocity computed by SNEC (indigo line) compared to the observed photospheric velocity (symbols) taken to be the velocity of the Si II ($\rm{\lambda}$ 6355) feature at early times (black squares) and Fe II ($\rm{\lambda}$5169) feature at later times (pink circles). 
The model deviates from the observations around the time that the photosphere begins to recede into the ejecta.
The model velocity at $\rm{\tau_{sob}=1}$ is plotted in yellow. 
While this provides a better agreement with the observations, the model velocity evolution is still faster than the observed evolution.
}
\label{fig:SNECVelocityCompare}
\end{center}
\end{figure}

Towards a consistent picture of the SN explosion, we also verified that the parameters we found from the light curve modeling were in agreement with the interpretation of other observables.
The high velocity derived from the spectroscopy is consistent with the large mass and explosion energy found by the light curve fitting.
It is also compatible with the incomplete gamma ray trapping, which is expected in SNe with high core velocities \citep{2011jerkstrand}.
Figure \ref{fig:SNECVelocityCompare} shows a comparison of the photospheric velocity of the best fit SNEC model to the observed photospheric velocity of ASASSN-15oz.
Following \citet{2014faran}, we define the photospheric velocity as that of Fe II ($\rm{\lambda}$5169). 
We use cachtio feature, now identified as Si II ($\rm{\lambda}$ 6355), to trace the photosphere at early times.
We note that while the Fe II ($\rm{\lambda}$5169) is often used to characterize the photospheric velocity, it is not well understood where in the atmosphere line is being formed. 
\citet{2018paxton} find that the Fe II ($\rm{\lambda}$5169) line is originating above the photosphere at a specific value of Sobolev optical depth ($\rm{\tau_{sob}}$) (they find $\rm{\tau_{sob} \sim1}$ fits the observations well).
However, \citet{2001hamuy}, in comparing the velocity derived from the Fe II ($\rm{\lambda}$5169) line to the photospheric velocity in the Type II SN atmospheric models of \citet{1996eastman}, find a systematic offset for individual objects that is in different directions for different objects, leading to no average offset. 
While there is fair agreement between the our model and observe velocity at early times, the model velocity is significantly lower than the observed velocity starting around day 25.
Following \citet{2018paxton}, we adjust our model velocity to the value at $\rm{\tau_{sob} = 1.0}$. 
Figure \ref{fig:SNECVelocityCompare} shows that this adjustment brings the SNEC velocities closer to the observed velocities.
A more detailed discussion concerning the modeling of the minimum of the FeII line is beyond the scope of this paper.

\begin{figure}
\begin{center}
\includegraphics[width=\columnwidth]{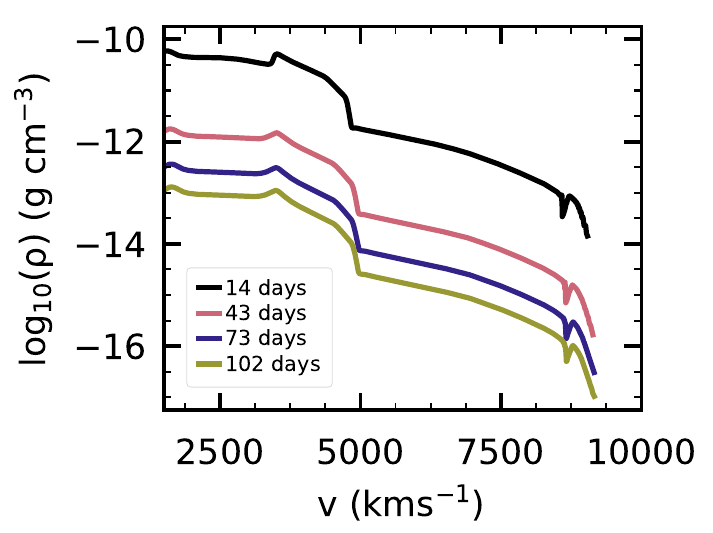} 
\caption{Density as a function of velocity in our best fit SNEC model at the free expansion stage (numbers in the plot indicate the time since explosion in days). 
The interface between the RSG envelope and the CSM corresponds to the velocity value of $\approx 4900\,{\rm km}\,{\rm s}^{-1}$. 
Mild increase in density at $\approx 3500\,{\rm km}\,{\rm s}^{-1}$ is a result of the reverse shock wave reflected from this interface.}
\label{fig:CDS}
\end{center}
\end{figure}

According to the classical model by \citet{1982chevalier}, the interaction of the SN ejecta with the surrounding wind leads to the formation of a cool dense shell between the forward and the reverse propagating shock waves. 
This shell is responsible for the X-ray emission from Type II SNe, and it may explain the broad base of the narrow emission lines seen in some interacting SNe \citep{2001chugai}. 
In our SNEC model, the CSM density is too high and the density drop between the envelope and the CSM is not sufficient to form a significantly overdense shell. 
This is supported by Figure \ref{fig:CDS}, where the interface between the RSG envelope and the CSM corresponds to the velocity coordinate of $\approx 4900\,{\rm km}\,{\rm s}^{-1}$. 
In this model, the forward shock wave sweeps the CSM almost entirely before the breakout.

On the other hand, if we were to consider a three component model, consisting of a RSG, a dense CSM and a regular low density stellar wind, like in \citet{2018morozova}, the thin cool dense shell would be formed at the interface between the dense CSM and the wind. 
In this scenario, the forward shock wave keeps propagating into the low density wind and generates radio emission. Since SNEC is not yet capable of handling the low density stellar wind, we have omitted it in our current simulations. 
However, we already know from the radio observations (Section \ref{sec:RadioInteraction}) that the mass-loss rate of this wind in ASASSN-15oz is $\approx 0.9-9\times 10^{-7}\,M_{\odot}\,{\rm yr}^{-1}$.
Such low density wind would not have any significant influence on the optical light curve of the SN, which justifies using a two component model (RSG and dense CSM only) for the purpose of this section.
\section{Discussion} \label{SecComp}
It is useful to compare observations of ASASSN-15oz to those of other IIP/IIL SNe to understand if it is a typical IIL-like SN or a unique event.
Although they are part of a continuous class, IIL-like SNe typically have higher velocity ejecta, are brighter, and have steeper light curve slopes than IIP-like SN.
In Section \ref{OpticalEvolve}, we showed that ASASSN-15oz has above average ejecta velocity.
Here we examine the light curve slope and absolute magnitude in the context of a sample of IIP/IIL SNe.

\begin{figure*}
\begin{center}
\includegraphics[width=\textwidth]{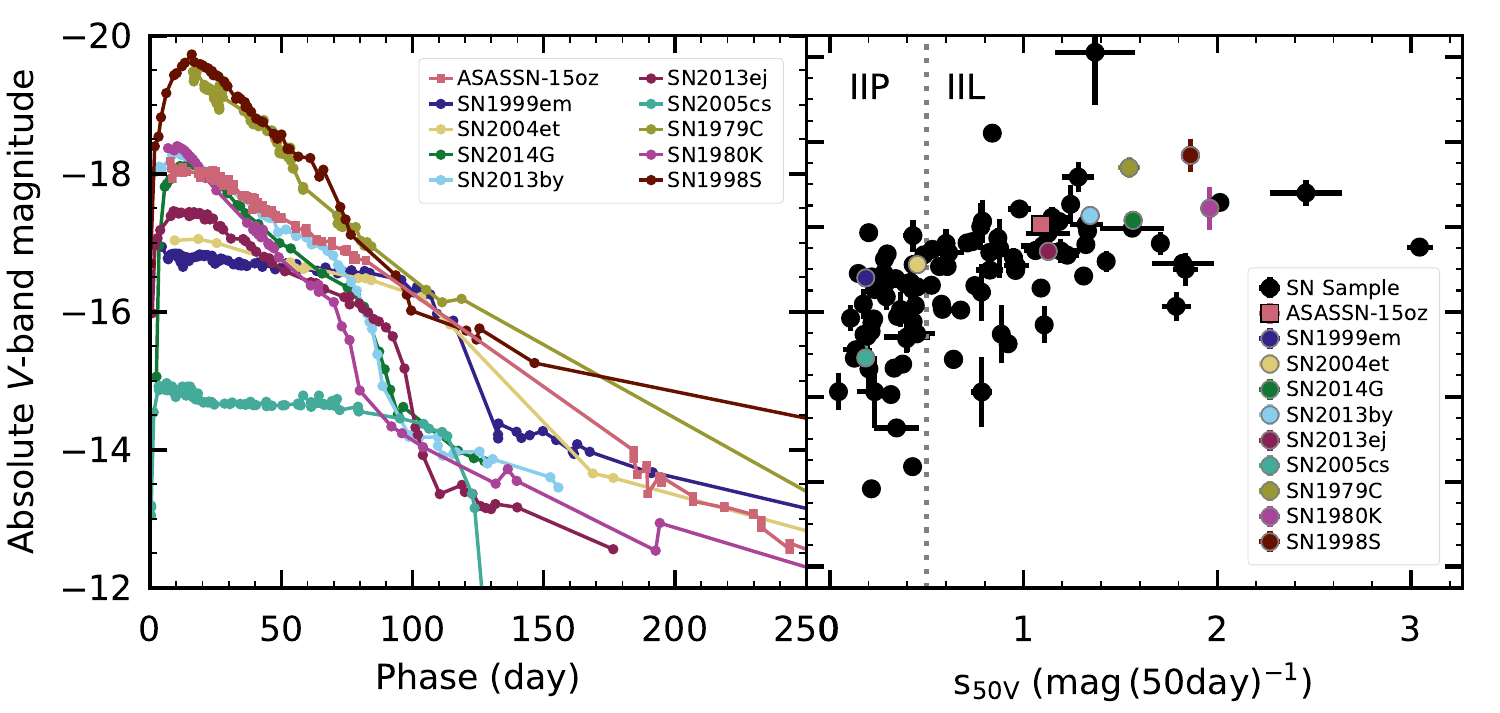} 
\caption{Left: the absolute {\it V}-band magnitude of ASASSN-15oz and nine well studied SNe that span the IIP/IIL like decline rates 
(1979C:  \citet{1980balinskaia,1981devaucouleurs,1982barbon},
1980K: \citet{1982barbon,1982buta,1983tsvetkov},
1998S: \citet{2000fassia,2000liu,2011li,2004pozzo},
1999em: \citet{2003elmhamdi},
2004et: \citet{2010maguire},
2005cs: \citet{2009pastorello},
2013by: \citet{2015valenti},
2013ej:  \citet{2014valenti},
2014G: \citet{2016terreran}). 
ASASSN-15oz is both bright and steeply declining when compared with this sample of classic SNe.
Also visible in this plot, is the steeper decline of the radioactive decay tail of ASASSN-15oz when compared with other objects.
Right: the absolute {\it V}-band magnitude compared to $\rm{s_{50V}}$ for a sample of public SNe from the SNDAVIS database (black circles). 
The trend seen by \citet{2014anderson} and \citet{2016valenti} of brighter SNe to have steeper slopes is apparent. 
ASASSN-15oz (pink square) follows this relationship. 
The SNe whose light curves are shown on the left are also plotted on the right for reference.
The gray dashed line denotes the separation between IIP-like and IIL-like SNe.
This separation is poorly defined in the literature.
The value we adopt is similar to that used by \citet{2014faran} and  \citet{2011li}.
}
\label{fig:SlopeMag}
\end{center}
\end{figure*}

Population studies of IIP/IIL SNe have found that SNe with steeper light curves are brighter \citep{2011li,2014anderson,2015sanders,2016valenti}, due to a larger explosion energy, a larger progenitor, CSM interaction, or a combined effect. 
The left panel of Figure \ref{fig:SlopeMag} shows the {\it V}-band light curve of ASASSN-15oz compared to nine other well studied SNe with a range of $\rm{s_{50V}}$.
ASASSN-15oz is brighter and more steeply declining than the classic IIP-like SNe 1999em and 2014et.
We also note that the slope of the radioactive tail of ASASSN-15oz is steeper than other objects plotted here.

We use our database of publicly available light curves, SNDAVIS, to compare ASASSN-15oz to a larger sample of objects.
In the right panel of Figure \ref{fig:SlopeMag} we plot the $\rm{s_{50V}}$ and the absolute {\it V}-band magnitude for 105 SNe. 
As a bright IIL-like SN, ASASSN-15oz is consistent with the sample correlation.

\begin{figure}
\begin{center}
\includegraphics[width=\columnwidth]{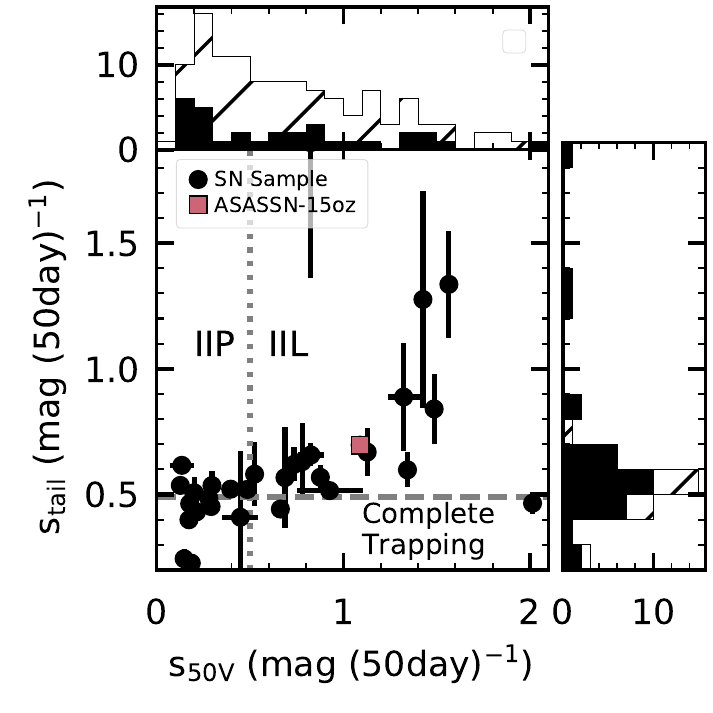} 
\caption{Center: the slope of the light curve at late times, during the radioactive decay phase compared to $\rm{s_{50V}}$ for a sample of publicly available SN light curves.
ASASSN-15oz is plotted in pink.
Top (Right): A histogram of the radioactive decay (photometric) phase slopes. All measured slopes from this phase are included in the hatched sample.
The objects for which there is both a decay phase slope and $\rm{s_{50V}}$  are plotted in black. The comparison of the full slope sample (hatch) to the combined slope sample (black) gives us confidence that points we plot in the center are representative of the full sample. 
The dashed grey line marks the radioactive tail slope corresponding to complete trapping with a slope of $\rm{s_{tail} = 0.5}$ mag (50 day)$\rm{^{-1}}$.
The dotted grey line represents the approximate separation between IIP and IIL SNe with a slope of $\rm{s_{50V}=0.5}$ mag (50 day)$\rm{^{-1}}$. 
 }
\label{fig:SlopeComp}
\end{center}
\end{figure}

We also compare $\rm{s_{50V}}$ and the radioactive decay slope  for ASASSN-15oz with all SNe in our database for which both slopes are measured in Figure \ref{fig:SlopeComp}.
The steepness of $\rm{s_{50V}}$ in a IIP/IIL SN light curve can be explained by the hydrogen recombination front receding faster than the ejecta expansion, implying a progenitor with a smaller hydrogen envelope.
The slope of the radioactive decay tail is determined by the fraction of gamma rays trapped and reprocessed by this same hydrogen envelope.
A lower fraction of trapped gamma rays can result either from a smaller hydrogen envelope or by external mixing of $\rm{^{56}Ni}$.
Given the possible dependence of both  $\rm{s_{50V}}$  and the radioactive decay slope on the hydrogen envelope, the correlation between them, seen in Figure \ref{fig:SlopeComp}, is unsurprising. 
ASASSN-15oz, having relatively steep slopes at both phases, lies at the IIL end of this relationship.
This correlation was presented in \citet{2014anderson} and is supported here with a larger sample.

Having analyzed individual observations for CSM interaction, we now turn to finding a unified model.
We derived mass-loss parameters from both the light curve modeling with SNEC and the modeling of radio observations.
From the optical light curve we found a compact CSM with a density of $\rm{K = 4.0\times10^{18}}$ g $\rm{cm^{-1}}$ extending to a radius of R = 1800 \rsun. 
Assuming a typical of RSG wind velocity between 10-100 km $\rm{s^{-1}}$, the CSM density required to fit the optical component implies a mass-loss rate between 0.8-8.0 \msunperiod $\rm{yr^{-1}}$.
This is significantly higher than radio mass-loss rate, $\rm{10^{-6}-10^{-7}}$ \msunperiod $\rm{yr^{-1}}$.
While the mass-loss rate from the radio observations is consistent with typical RSG mass-loss rates of $\rm{10^{-4} }$- $\rm{10^{-6}}$ \msunperiod $\rm{yr^{-1}}$, the optical mass-loss rate is consistent with some late stage mass-loss scenarios (e.g. \citealt{2001chugai,2012quatert,2014shiode,2017fuller, 2017yaron}).
These discrepant values can be reconciled by considering the origin of the radiation we are observing.
The excess luminosity seen in the optical light curve is due to radiation diffusion from the shocked CSM that is behind the forward shock and is only visible after shock breakout.
The radio luminosity on the other hand, is created by the acceleration of electrons due to the shock's interaction with the CSM.
It is therefore representative of the shock's location at the time of the observation.
Our observations can then be explained by a long period of average mass-loss, followed by a short period of extreme mass-loss.

\begin{figure}
\begin{center}
\includegraphics[width=\columnwidth]{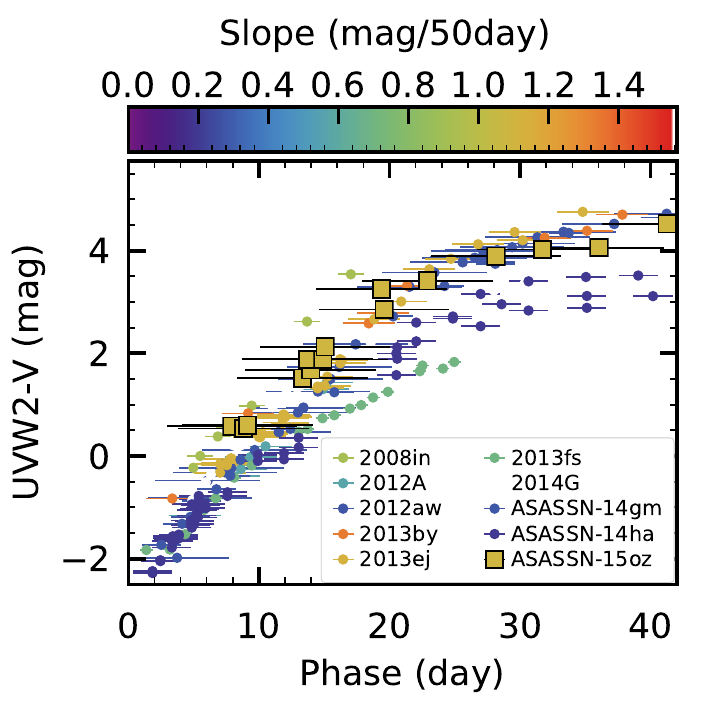} 
\caption{A comparison of the UV color of ASASSN-15oz to that of all other IIP/IIL SNe with \textit{Swift} observations and $\rm{s_{50V}}$ measurements in SNDAVIS.
Each point is colored by the $\rm{s_{50V}}$ slope, with shallower slopes in blue and steeper slopes in red. 
Based on this data, there is no relationship between UV excess and slope.}
\label{fig:UVColor}
\end{center}
\end{figure}

With this model in mind, we examine our UV and X-ray observations, in the context of other IIP/IIL SNe.
Strongly interacting SNe remain UV bright throughout their interaction, implying that perhaps a UV excess would be observed for IIP/IIL SNe during the first 30 days, when the CSM has the greatest affect on the light curve.
Furthermore, if in fact, the diversity in light curve shapes is due to mass-loss and the interaction is strong enough to produce a UV excess, then those SNe with a steeper slope, should have lost more mass, and therefore should be more UV bright.
We explore this by plotting {\it Swift} UV color, UVW2 - V, over the first 30 days of evolution for eight SNe in our database with {\it Swift} UVW2 and {\it V}-band photometry and $\rm{s_{50V}}$ measurements (Figure \ref{fig:UVColor}). 
The symbols for each SN are colored by the {\it V}-band slope at 50 days. 
Although this sample is small, a wide range of slopes are represented and no correlation between UV excess and slope is observed.
We interpret this as an indication that the CSM interaction is not strong enough to produce UV excess.
However, we cannot rule out the possibility that a correlation is disrupted by uncertainties in the extinction correction.

\begin{figure}
\begin{center}
\includegraphics[width=\columnwidth]{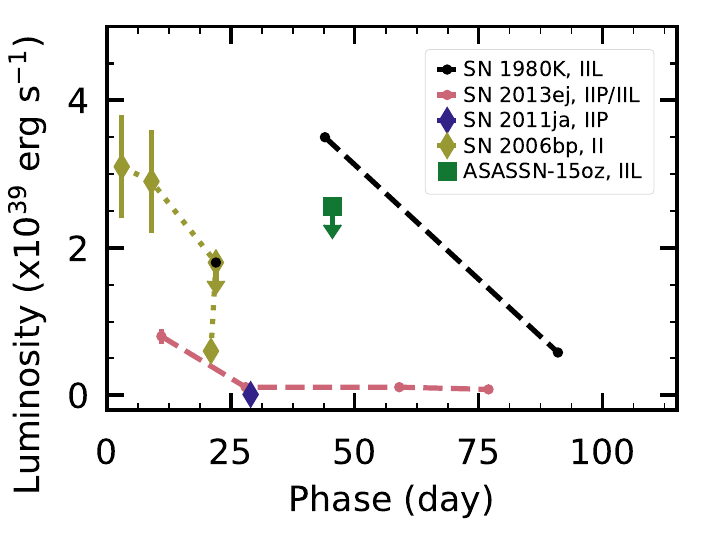} 
\caption{The X-ray luminosity of all published SNe found in the SNaX database. 
IIP-like SNe are shown with diamond symbols while IIL-like SNe are shown with circular symbols and the upper limit of ASASSN-15oz is marked with a square.
Many detections from IIP/IIL SNe are below the upper limit derived in this paper, implying that deeper observations would be required to detect its X-ray flux.}
\label{fig:xray}
\end{center}
\end{figure}

Like the UV excess, X-rays from strongly interacting SNe are frequently observed.
To gain more insight into the X-ray upper limit, we compare the upper limit of ASASSN-15oz with published X-ray observations of IIP/IIL SNe found in the SNaX database \citep{2017ross} (see Figure \ref{fig:xray}).
We find our upper limit is above some other detections and upper limits of IIP/IIL SNe in the database.
This further suggests that deeper observations could be needed to detect the X-ray flux of ASASSN-15oz.

\section{Conclusions}\label{sec:conclude}
We present extensive multi-wavelength observations of the type IIL-like SN, ASASSN-15oz, including two epochs of radio observations, four epochs of NIR spectroscopy, three epochs of NIR photometry, a 400 day dense multi-band optical light curve, dense optical spectroscopy during the photospheric phase, three epochs of nebular optical spectroscopy,  one epoch of UV spectroscopy, and 60 day UV and X-ray light curve.

Optical and NIR observations from the photospheric phase show ASASSN-15oz to be a typical IIL-like SN, with a linearly declining light curve and above average photospheric velocities. 
We find that in the explosion of ASASSN-15oz 0.08 -0.11 \msun of $\rm{{}^{56}Ni}$ was synthesized. 
However, since both the light curve modeling and the nebular spectra point towards the lower limit, we consider the lower limit most likely.
Its bright peak magnitude is consistent with the trend of increasing brightness with increasing photospheric slope. 
The steep decline of the late time light curve is also congruous with the direct correlation found between $\rm{s_{50V}}$ and radioactive tail slopes.
Optical light curve modeling using SNEC finds an 17 \msun progenitor, consistent with the range of progenitors derived from modeling of the nebular spectra.

We search for signs of interaction at all wavelengths with the following results:
\begin{itemize}
\item We detect the ASASSN-15oz in radio observations, implying a mass-loss rate of $\rm{0.9-9\times10^{-7}}$ \msunperiod $\rm{yr^{-1}}$ for a wind velocity between 10-100 km $\rm{s^{-1}}$.
\item High velocity H and He are searched for in the NIR and optical spectra. 
Although there is a feature bluewards of H $\rm{\alpha}$, no corresponding feature is observed in any other hydrogen line nor in the He I ($\rm{\lambda 10830}$) line. 
For this reason we identify the cachito feature as Si II ($\rm{\lambda 6355}$) rather than high velocity hydrogen.
\item To accurately model the early optical light curve with an evolutionary progenitor, we require CSM around the RSG progenitor with a density of $\rm{4\times10^{18}}$ g $\rm{cm^{-1}}$ and radial extent of 1800 $\rm{R_{\odot}}$, implying a short period of extreme mass-loss.
\item No narrow emission lines are detected in the optical and UV photospheric spectroscopy, implying a mass-loss rate of $\rm{< 10^{-4}}$ \msunperiod $\rm{yr^{-1}}$.
\item The H $\rm{\alpha}$ emission line profile in the nebular spectra is asymmetric, which could be the result of interaction with CSM.
\item No UV excess is detected in the first 40 days of observations.
\item No X-ray signal is detected in the sum of all photospheric observations, implying $\rm{L_{xray}<2.56 \times10^{39}}$ erg $\rm{s^{-1}}$.
\end{itemize}
Taking all of these results together, we find the progenitor of ASASSN-15oz underwent standard RSG mass-loss for most of its evolution, with a short period of extreme mass-loss just before explosion. 
It is possible that early optical spectroscopy would have shown narrow emission lines as the shock illuminated the edge of the dense shell of material.
Additionally, an X-ray detection may have been made with deeper observations. 
A dense shell of material surrounded by a lower density CSM has now been used to model the observations of a handful of objects (e.g. 1998S: \citealt{2001chugai}, 2013fs: \citealt{2017yaron}; 2013ej: \citealt{2018morozova2}).
As the best model to explain the observations of a IIP-like SN (2013fs), a transition SN (2013ej), and now to a IIL-like SN (ASASSN-15oz) it may be applicable to a broad range of progenitors.

\section*{Acknowlegments}
This work was supported by the {\it Swift} Guest Observers Program through grant NNX16AE90G.
This work is based (in part) on observations collected at the European Organization for Astronomical Research in the Southern Hemisphere, Chile as part of PESSTO, (the Public ESO Spectroscopic Survey for Transient Objects Survey) ESO program 188.D-3003, 191.D-0935, 197.D-1075.
Based on observations collected at the European Organization for Astronomical Research in the Southern Hemisphere under ESO program 095.A-0316(A). 
We acknowledge support from EU/FP7-ERC grant [615929].
DAH, GH and CM are supported by NSF grant AST-1313484.
KML acknowledges funding from the European Research Council under ERC Consolidator Grant agreement no 647208.
A.L.P. acknowledges financial support for this research from a Scialog award made by the Research Corporation for Science Advancement.
O.R. acknowledge support by projects IC120009 ``Millennium Institute of Astrophysics (MAS)'' of the Iniciativa Cient\'ifica Milenio del Ministerio Econom\'ia, Fomento y Turismo de Chile and CONICYT PAI/INDUSTRIA 79090016. 
DJS is a visiting astronomer at the Infrared Telescope Facility, which is operated by the University of Hawaii under contract NNH14CK55B with the National Aeronautics and Space Administration.
Research by DJS is supported by NSF grants AST-1821967, 1821987, 1813708 and 1813466.
This research has made use of the NASA/IPAC Extragalactic Database (NED), which is operated by the Jet Propulsion Laboratory, California Institute of Technology, under contract with the National Aeronautics and Space Administration.
This work uses the National Radio Astronomy Observatory which is a facility of the National Science Foundation operated under cooperative agreement by Associated Universities, Inc.
This work makes use of observations from the LCO network. 
This work was partly supported by the UK Space Agency.
Based on observations obtained at the Gemini Observatory through proposal GS-2016A-Q-75-25, acquired through the Gemini Observatory Archive, and processed using the Gemini IRAF package, which is operated by the Association of Universities for Research in Astronomy, Inc., under a cooperative agreement with the NSF on behalf of the Gemini partnership: the National Science Foundation (United States), the National Research Council (Canada), CONICYT (Chile), Ministerio de Ciencia, Tecnolog\'{i}a e Innovaci\'{o}n Productiva (Argentina), and Minist\'{e}rio da Ci\^{e}ncia, Tecnologia e Inova\c{c}\~{a}o (Brazil).
This research made use of Astropy,\footnote{http://www.astropy.org} a community-developed core Python package for Astronomy \citep{astropy:2013, astropy:2018}
Support for G.P. is provided by the Ministry of Economy, Development, and Tourism's Millennium Science Initiative through grant IC120009, awarded to The Millennium Institute of Astrophysics, MAS.\\

\bibliographystyle{mnras}
\bibliography{references}

\appendix
\section{Tables of Observations}
\begin{table*}
\caption{Spectroscopic Observations of ASASSN-15oz.\label{tab:SpecObs}}
\begin{tabular}{ccccc}
\hline
Date & JD & Phase (Day) & Observatory & Instrument \\
\hline
2015-09-04 & 2457270.0 & 8.0 & LCO & FLOYDS \\
2015-09-05 & 2457270.7 & 8.7 & Swift & UVOTA \\
2015-09-05 & 2457271.1 & 9.1 & Swift & UVOTA \\
2015-09-05 & 2457270.6 & 8.6 & NTT & SOFI \\
2015-09-06 & 2457272.0 & 10.0 & LCO & FLOYDS \\
2015-09-07 & 2457273.0 & 11.0 & LCO & FLOYDS \\
2015-09-10 & 2457275.7 & 13.7 & Swift & UVOTA \\
2015-09-10 & 2457275.7 & 13.7 & Swift & UVOTA \\
2015-09-11 & 2457277.0 & 15.0 & LCO & FLOYDS \\
2015-09-11 & 2457276.9 & 14.9 & Swift & UVOTA \\
2015-09-16 & 2457282.0 & 20.0 & LCO & FLOYDS \\
2015-09-20 & 2457286.1 & 24.1 & LCO & FLOYDS \\
2015-09-21 & 2457287.5 & 25.5 & VLT & X-SHOOTER \\
2015-09-24 & 2457290.0 & 28.0 & LCO & FLOYDS \\
2015-09-30 & 2457296.0 & 34.0 & LCO & FLOYDS \\
2015-10-04 & 2457299.5 & 37.5 & NTT & EFOSC \\
2015-10-05 & 2457300.5 & 38.5 & NTT & SOFI \\
2015-10-06 & 2457301.9 & 39.9 & LCO & FLOYDS \\
2015-10-10 & 2457305.7 & 43.7 & IRTF & SpeX \\
2015-10-14 & 2457310.0 & 48.0 & LCO & FLOYDS \\
2015-10-25 & 2457320.9 & 58.9 & LCO & FLOYDS \\
2015-11-07 & 2457333.9 & 71.9 & LCO & FLOYDS \\
2015-11-08 & 2457334.5 & 72.5 & NTT & EFOSC \\
2015-11-19 & 2457345.5 & 83.5 & NTT & EFOSC \\
2016-04-11 & 2457489.9 & 227.9 & NTT & EFOSC \\
2016-04-11 & 2457489.9 & 227.9 & NTT & EFOSC \\
2016-06-09 & 2457548.8 & 286.8 & Gemini-S & GMOS \\
2016-06-10 & 2457549.7 & 287.7 & Gemini-S & GMOS \\
2016-06-12 & 2457551.7 & 289.7 & Gemini-S & GMOS \\
2016-08-03 & 2457603.7 & 341.7 & NTT & EFOSC \\
2016-09-11$\rm{^{*}}$ & 2457642.6 & 380.6 & NTT & EFOSC \\
2016-09-11$\rm{^{*}}$ & 2457642.6 & 380.6 & NTT & EFOSC \\
2016-09-19 & 2457650.5 & 388.5 & NTT & EFOSC \\
2017-09-20$\rm{^{+}}$ & 2458016.9 & 754.9 & Swift & UVOTA \\
\hline
\end{tabular}
\\$\rm{^{*}}$No signal in data due to cloud cover\\ $\rm{^{+}}$ Template observation
\end{table*}
\begin{table*}
\centering
\caption{Sample of Imaging Observations of ASASSN-15oz. Full table available on-line.\label{tab:LcObs}}
\begin{tabular}{ccccccc}
Date-Obs & JD & Phase (Day) & Apparent Magnitude & Apparent Magnitude Error & Filter & Source \\
\hline
2015-09-04 & 2457269.68 & 7.68 & 14.64 & 0.02 & B & LSC 1m \\
2015-09-04 & 2457269.68 & 7.68 & 14.66 & 0.02 & B & LSC 1m \\
2015-09-04 & 2457269.69 & 7.69 & 14.37 & 0.02 & V & LSC 1m \\
2015-09-04 & 2457269.69 & 7.69 & 14.43 & 0.01 & g & LSC 1m \\
2015-09-04 & 2457269.69 & 7.69 & 14.42 & 0.01 & g & LSC 1m \\
\end{tabular}
\end{table*}
\section{Line Identification and Fitting Details}\label{AppLineFit}
\begin{table*}
\caption{The best parameters for the SYN++ fit. 
The Sobolev opacity $\rm{\tau}$ is modeled with an exponential profile with e-folding length, aux, minimum velocity, $\rm{v_{min}}$, and maximum velocity, $\rm{v_{max}}$. 
The temperature column is the Boltzmann excitation temperature.
SYN++ models pure reasonance scattering which is a poor approximation for the H $\rm{\alpha}$. 
For this reason we use the H $\rm{\beta}$ line to determine the hydrogen contribution to the fit.
A separate fit is performed for H $\rm{\alpha}$ and is listed separately in the table.
We emphasize that this fit is used for line identification and not to derive ejecta properties.}
\begin{center}
\begin{tabular}{c|c|c|c|c|c}
\hline
Ion & log($\rm{\tau}$) & $\rm{v_{min}(kkm}$ $\rm{s^{-1})}$ & $\rm{v_{max}(kkm/s)}$ & aux & Temperature (kK) \\
\hline
  H I -$\rm{\alpha}$   & 0.06 & 0.1 & 40.0 & 2.0 & 10.0 \\
  H I - $\rm{\beta}$   & 1.1& 0.1 & 40.0 & 2.0 & 10.0 \\
  Na II & -0.5 & 0.1 & 40.0 & 1.0 & 10.0 \\
  O I & -0.8 & 0.1 & 40.0 & 1.0 & 10.0 \\
  Ca II & 1.3 & 0.1 & 40.0 & 2.0 & 10.0 \\
  Sc II & 0.1 & 0.1 & 40.0 & 1.0 & 10.0 \\
  Ti II  & 0.3 & 0.1 & 40.0 & 1.0 & 10.0 \\
  Fe I  &  0.3 & 0.1 & 40.0 & 1.0 & 10.0 \\
  Fe II & 0.3 & 0.1 & 40.0 & 1.0 & 7.0 \\
  Ba I & 0.0 & 0.1 & 40.0 & 1.0 & 10.0 \\  
  \hline
 \end{tabular}
\end{center}
\label{tab:syn++}
\end{table*}

\begin{figure*}
\begin{center}
\includegraphics[width=\textwidth]{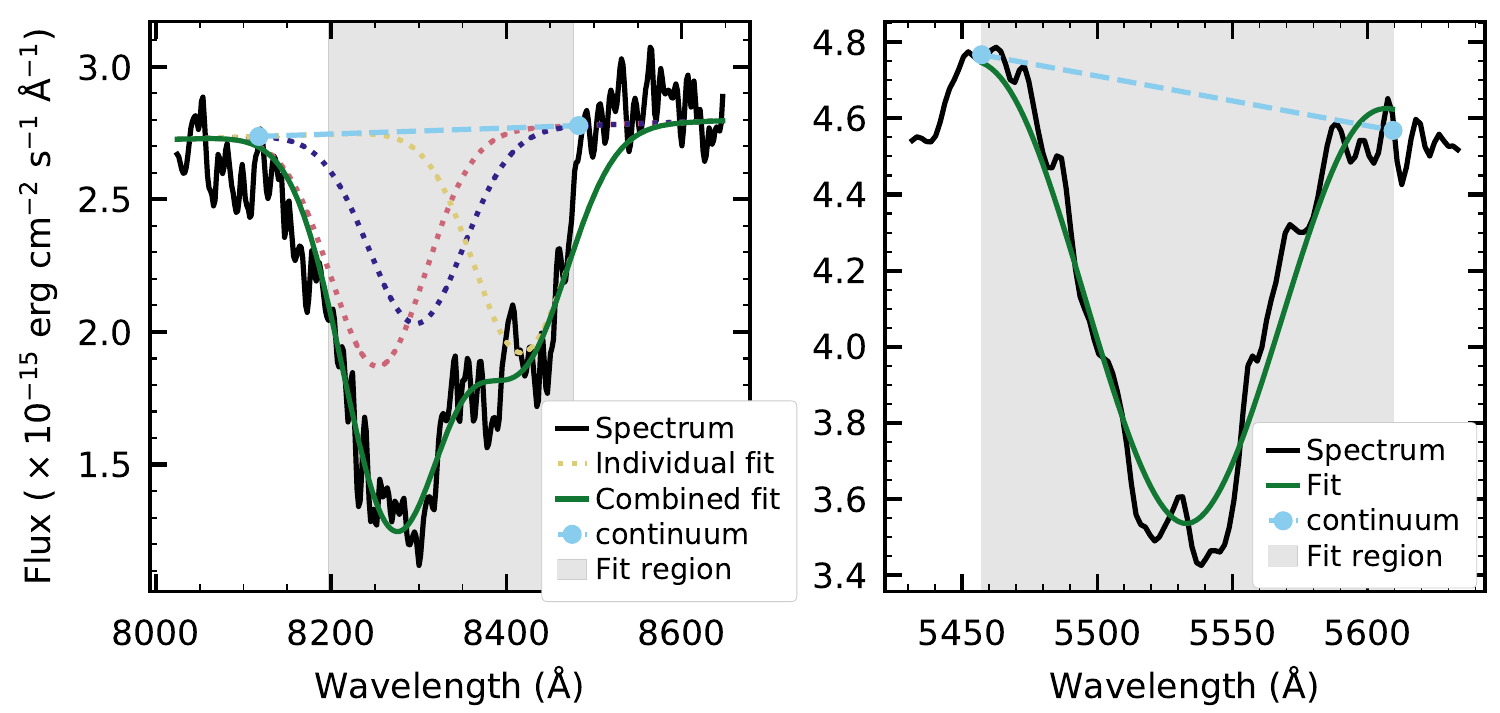}
\caption{An example of a fit to the multi-component Ca II NIR triplet (left) using multiple Gaussians and a single line fit (following \citealt{2012silverman}, right) to the Sc II ($\rm{\lambda}$5662) line.
The observed spectrum is plotted in black.
The continuum edges are marked with a cyan circle and the cyan dashed line connecting these points is used as the continuum. 
The Ca II NIR triplet fit is found by simultaneously fitting 3 Gaussians with the same standard deviation and mean offsets corresponding to the expected wavelength separation of the triplet. 
The individual Gaussians are plotting at as dotted lines and the combined fit is plotted as a solid green line. 
The minima of the individual Gaussians is used to find the velocity of each component.
The Sc II feature is fit with a cubic spliine. 
The minimum of the spline is used to find the Sc II velocity.
\label{fig:VelocityFit}}
\end{center}
\end{figure*}

For lines that are blended (but still showed partially isolated absorption troughs for some features) we define the continuum as a straight line from the lowest side of the feature. 
After dividing by the continuum, we then simultaneously fit multiple 1D Gaussian profiles to the continuum normalized spectra. 
For blended features originating in the same part of the ejecta, we require that each component has the same width.
We define the velocity of a feature as the minimum (mean) of the individual Gaussian profiles corresponding to that ion and the error as the standard deviation of the fit, representing the range of velocities present in the photosphere. 
An example of the multiple components and the combined fit is shown in the left panel of Figure \ref{fig:VelocityFit}.
For the Ca II NIR triplet we use the additional constraint of a fixed offset between the minima of the features. 

For unblended lines, we perform the following steps to find the minimum and the standard deviation (following \citealt{2012silverman}).
We eliminate large noise spikes (e.g. from cosmic rays) using a Savitzky-Golay smoothing filter \citep{1964savitzky} with a quadratic function over a binsize of five pixels. 
For each line, a minimum and maximum wavelength is defined as well as a slope to account for the shifting of the feature over time as the ejecta slows. 
A slope is then fit to small bins over this wavelength range, starting with a binsize of five pixels and increasing until exactly three changes in slope are found (at either edge of the line and the line center). 
If the binsize reaches 40 percent of the feature size, no further attempt at fitting is made. 
The feature edges are confirmed by fitting a quadratic function to the region centered on each edge identified by the slope change with a width of 20 pixels for the FLOYDS spectra and five pixels for the EFOSC2 spectra. 
The edge is considered successfully found if the quadratic fit is concave down. 
The continuum is defined by a line though these end points.
Finally, the velocity is found by fitting a cubic spline to the continuum subtracted flux between the edges using the flux errors as weights. 
An example of a feature with the continuum and spline fit is shown in the right panel of Figure \ref{fig:VelocityFit}.

We use both methods to fit Na I ($\rm{\lambda}$ 5898) and H $\rm{\beta}$ ($\rm{\lambda}$4861). 
For Na I, we find the spline fit does not characterize the minimum of the profile well and instead use the Gaussian minimum. 
The H $\rm{\beta}$ profile is contaminated by another feature (possibly Ti II) at later times, offsetting the minimum found using the spline fit. 
For this reason we prefer the Gaussian fit for H $\rm{\beta}$ as well. 
A comparison of the results of both fits for H $\rm{\beta}$ at early times finds them in good agreement, giving us confidence in the consistency of the two methods.

\bsp	
\label{lastpage}
\end{document}